\documentclass[amsmath,amssymb]{revtex4}
\usepackage{graphicx}
\usepackage{amscd,amsmath,amsthm,amsfonts,amssymb}

\def\[{\begin{equation}}
\def\]{\end{equation}}

\begin{document}
\title{Rogue waves in the generalized derivative nonlinear Schr\"{o}dinger equations}
\author{Bo Yang$^a$, Junchao Chen$^b$ and Jianke Yang$^a$}
\affiliation{$^a$Department of Mathematics and Statistics, University of Vermont, Burlington, VT 05405, U.S.A. \\
$^b$Department of Mathematics, Lishui University, Lishui 323000, China}

\begin{abstract}
General rogue waves are derived for the generalized derivative nonlinear Schr\"odinger (GDNLS) equations by a bilinear Kadomtsev-Petviashvili (KP) reduction method. These GDNLS equations contain the Kaup-Newell equation, the Chen-Lee-Liu equation and the Gerdjikov-Ivanov equation as special cases. In this bilinear framework, it is shown that rogue waves to all members of these equations
are expressed by the same bilinear solution. Compared to previous bilinear KP reduction methods for rogue waves in other integrable equations, an important improvement in our current KP reduction procedure is a new parameterization of internal parameters in rogue waves. Under this new parameterization, the rogue wave expressions through elementary Schur polynomials are much simpler. In addition, the rogue wave with the highest peak amplitude at each order can be obtained by setting all those internal parameters to zero, and this maximum peak amplitude at order $N$ turns out to be $2N+1$ times the background amplitude, independent of the individual GDNLS equation and the background wavenumber.
It is also reported that these GDNLS equations can be decomposed into two different bilinear systems which require different KP reductions, but the resulting rogue waves remain the same. Dynamics of rogue waves in the GDNLS equations is also analyzed. It is shown that the wavenumber of the constant background strongly affects the orientation and duration of the rogue wave. In addition, some new rogue patterns are presented.
\end{abstract}

\maketitle

\section{Introduction}
Rogue waves are large and spontaneous local excitations of nonlinear wave equations that ``appear from nowhere and
disappear with no trace" \cite{Akhmediev_2009}. More specifically, these local excitations arise from a flat constant-amplitude background, reach a transient high amplitude, and then retreat back to the same flat background. Such solutions were first reported for the nonlinear Schr\"{o}dinger (NLS) equation by Peregrine in 1983 \cite{Peregrine}. In recent years, such waves were linked to freak waves in the ocean \cite{Ocean_rogue_review,Pelinovsky_book} and extreme events in optics \cite{Solli_Nature,Wabnitz_book}, and were observed in water-tank and optical-fiber experiments \cite{Tank1,Tank2,Fiber1,Fiber2,Fiber3}. Motivated by these physical applications, rogue waves have been derived in a large number of physically-relevant integrable nonlinear wave equations, including the NLS equation \cite{AAS2009,DGKM2010,ACA2010,KAAN2011,GLML2012,OhtaJY2012,DPMVB2013}, the derivative NLS equations \cite{KN_rogue_2011,KN_rogue_2013,CCL_rogue_Chow_Grimshaw2014,CCL_rogue_2017}, the Manakov equations \cite{BDCW2012,ManakovDark}, the Davey-Stewartson equations \cite{OhtaJKY2012,OhtaJKY2013}, and many others \cite{AANJM2010,OhtaJKY2014,ASAN2010,Chow,MuQin2016,LLMFZ2016,ClarksonDowie2017,YangYang2019Nonloc,JCChen2018LS,XiaoeYong2018}.
Indeed, rogue waves are caused by baseband modulation instability of the constant-amplitude background \cite{ManakovDark}. Thus, any integrable equation with baseband modulation instability is expected to admit rogue waves, which can be derived by integrable techniques. All known rogue waves in integrable equations are rational solutions of the underlying systems. This fact is related to baseband modulation instability, since rational rogue-wave solutions are associated with long-wave instability of the background. We note by passing that in nonintegrable systems, large and spontaneous local excitations can also arise from a constant-amplitude background if such background admits baseband modulation instability (see \cite{Solli_Nature} for instance). But such excitations do not retreat back to the same background, and are not expected to admit exact analytical expressions, due to the lack of integrability of the underlying nonlinear wave equations \cite{Akhmediev_rogue_Raman}.

In this paper, we consider rogue waves in the generalized derivative nonlinear Schr\"{o}dinger (GDNLS) equations \cite{Kundu1984,Clarkson1987}
\[ \label{GDNLS0}
\textrm{i} \phi_{t} + \phi_{\xi\xi} + \rho |\phi|^2\phi + \textrm{i}a \phi\phi^*\phi_\xi + \textrm{i}b \phi^2\phi_\xi^* +\frac{1}{4}b(2b-a)|\phi|^4\phi=0,
\]
where $\rho, a, b$ are arbitrary real constants with $a\ne b$, and the superscript `*' represents complex conjugation (the $a=b$ case will be treated in the appendix). In fiber optics, these equations model the propagation of short light pulses where, in addition to dispersion and Kerr (cubic) nonlinearity, self-steepening and fifth-order nonlinearity are also accounted for (even though the Raman effect and third-order dispersion are omitted) \cite{Agrawal_book,Kivshar_book}. When $\rho=0$ and $b=2a$, these equations reduce to the Kaup-Newell equation \cite{Kaup_Newell}, which governs the propagation of circularly polarized nonlinear Alfv\'en waves in plasmas \cite{KN_Alfven1,KN_Alfven2}. When $\rho=b=0$, these equations reduce to the Chen-Lee-Liu equation \cite{CCL}, which models short-pulse propagation in a frequency-doubling crystal through the interplay of quadratic and cubic nonlinearities \cite{Wise_CCL}. Due to these physical applications, rogue wave formation in these GDNLS equations is a physically significant issue.

There have been a number of studies on rogue waves in these GDNLS equations. For instance, for the Kaup-Newell equation (with $\rho=0$ and $b=2a$), special types of rogue waves were derived by Darboux transformation in \cite{KN_rogue_2011,KN_rogue_2013}. For the Chen-Lee-Liu-type equation, with $b=0$ in (\ref{GDNLS0}), the fundamental rogue wave was derived by the bilinear Hirota method in \cite{CCL_rogue_Chow_Grimshaw2014}, and higher-order rogue waves were derived by Darboux transformation in \cite{CCL_rogue_2017}. For the Gerdjikov-Ivanov equation \cite{GI}, with $\rho=a=0$ in (\ref{GDNLS0}), fundamental and higher-order rogue waves were derived by Darboux transformation in \cite{GI_rogue_2012,GI_rogue_2014}. Even for the GDNLS equations (\ref{GDNLS0}) themselves, general rogue waves were derived by Darboux transformation in \cite{GDNLS_rogue2019}, and their chirping phase structure was examined.

In this article, we derive general rogue waves in the GDNLS equations (\ref{GDNLS0}) by the bilinear Kadomtsev-Petviashvili (KP) reduction method. The advantage of this bilinear framework is that rogue waves in all GDNLS equations (\ref{GDNLS0}) can be expressed explicitly by the same bilinear solution. Compared to previous bilinear KP reduction methods for rogue waves in other integrable equations
\cite{OhtaJY2012,OhtaJKY2012,OhtaJKY2013,OhtaJKY2014,JCChen2018LS,XiaoeYong2018,YangYang2019Nonloc}, an important improvement in our current KP reduction technique is a new parameterization of internal parameters in rogue waves. Under this parameterization,
analytical expressions of rogue waves through Schur polynomials are much simpler. More importantly, when all internal parameters are set to zero, we would get a parity-time-symmetric rogue wave which attains the maximum peak amplitude among rogue waves of that order. This allows us to analytically derive this maximum peak amplitude, which turns out to be $2N+1$ times the background amplitude at order $N$, independent of the individual GDNLS equation and the background wavenumber. We also find that the GDNLS equations (\ref{GDNLS0}) can be decomposed into two different bilinear systems which require different KP reductions, but the resulting rogue waves are the same. After these rogue waves are derived, their dynamics is also analyzed. It is shown that the wavenumber of the background strongly affects the orientation and duration of the rogue wave. In addition, some new rogue patterns are presented. In the appendix, general rogue waves for the GDNLS equations (\ref{GDNLS0}) with $a=b$ (the so-called Kundu-Eckhaus equation) are also given in the bilinear framework. These results deepen our understanding of rogue waves in the physically significant GDNLS equations (\ref{GDNLS0}). Meanwhile, they advance the bilinear KP-reduction methodology for the derivation of rogue waves.

\section{Preliminaries} \label{sec:pre}
Under a simple gauge transformation \cite{Satsuma_GDNLS_soliton}
\begin{equation*}
\phi(\xi, t)=\sqrt{\frac{2}{a-b}}\hspace{0.1cm} u(x, t)\hspace{0.1cm} \exp\left\{\textrm{i}\frac{\rho}{a-b}x+\textrm{i}\frac{\rho^2}{(a-b)^2}t\right\},
\end{equation*}
where $x=\xi-2\rho t/(a-b)$, the GDNLS equations (\ref{GDNLS0}) with $a\ne b$ reduce to
\[ \label{GDNLS}
\textrm{i}u_t+u_{xx}+2\textrm{i}\gamma |u|^2u_x+2\textrm{i}(\gamma-1)u^2u_x^*+(\gamma-1)(\gamma-2)|u|^4u=0,
\]
where $\gamma=a/(a-b)$. We will work with these normalized GDNLS equations (\ref{GDNLS}) in the remainder of this article. These equations become the Kaup-Newell equation when $\gamma=2$ \cite{Kaup_Newell}, the Chen-Lee-Liu equation when $\gamma=1$ \cite{CCL}, and the Gerdjikov-Ivanov equation when $\gamma=0$ \cite{GI}.

It is noted that with an additional gauge transformation
\[  \label{gauge}
u(x,t)=v(x,t)\hspace{0.05cm} e^{\textrm{i}(2-\gamma)\int |v(x,t)|^2 dx},
\]
the normalized GDNLS equations (\ref{GDNLS}) further reduce to the Kaup-Newell equation
\[ \label{KNE}
\textrm{i}v_t+v_{xx}+2\textrm{i}(|v|^2v)_{x}=0.
\]
Thus, from rogue waves of the Kaup-Newell equation, one can derive rogue waves in the GDNLS equations (\ref{GDNLS}) in principle. However, the gauge transformation (\ref{gauge}) involves a nontrivial integral, which makes it difficult to derive explicit solutions to the GDNLS equations from those of the Kaup-Newell equation. For this reason, we will not utilize this gauge transformation. Instead, we will use a bilinear method to directly obtain explicit rogue wave solutions in the GDNLS equations (\ref{GDNLS}) for arbitrary $\gamma$ values.

Rogue waves in the GDNLS equations (\ref{GDNLS}) approach a constant-amplitude continuous wave background at large $x$ and $t$. By simple variable scalings, this constant amplitude can be normalized to be unity. Then, these rogue waves approach the unit-amplitude continuous wave background $e^{\textrm{i} \kappa x -\textrm{i}\omega t}$, where $\kappa$ is a free wavenumber, and $\omega=\kappa^2+2\kappa-(\gamma-1)(\gamma-2)$ is the frequency. In order for rogue waves to arise, these backgrounds must be unstable to baseband modulations \cite{ManakovDark}. Simple modulation instability calculations show that these backgrounds are base-band unstable when $\kappa<1-\gamma$.
Thus, rogue waves in the GDNLS equations (\ref{GDNLS}) should approach the following background as $x, t \rightarrow \pm \infty$:
\begin{eqnarray}
&& u(x,t)\rightarrow  e^{{\rm i}(1-\gamma-\alpha) x-{\rm i}\left[\alpha^2+2(\gamma-2)\alpha+1-\gamma\right]t},  \label{BoundaryCond1}
\end{eqnarray}
where $\alpha>0$ is a wavenumber parameter. Unlike the NLS equation, the GDNLS equations (\ref{GDNLS}) do not admit Galilean-transformation invariance. Thus, $\alpha$ is a non-reducible parameter in its rogue waves.

In view of the above boundary condition, we introduce the variable transformation
\begin{eqnarray}
&& u= e^{{\rm i}(1-\gamma-\alpha) x-{\rm i}\left[\alpha^2+2(\gamma-2)\alpha+1-\gamma\right]t}\hspace{0.05cm}\frac{(f^*)^{\gamma-1}g}{f^\gamma}, \label{BilinearTrans}
\end{eqnarray}
where $f$ and $g$ are complex functions. Under this transformation, the GDNLS equations (\ref{GDNLS}) can be decomposed into the following system of four bilinear equations:
\begin{eqnarray}
&& \left(\mathrm{i}D_{t}+D_{x}^{2}+2\mathrm{i}(1-\alpha)D_{x} \right) g\cdot f^*=0, \label{dnlsbilinEq11}\\
&& \left({\rm i} D_t + D^2_x +2{\rm i}  D_x \right) f\cdot f^*=0, \label{dnlsbilinEq12}\\
&& \left(\mathrm{i} D_{x}-1\right) f\cdot f^{*}+ |g|^{2}=0, \label{dnlsbilinEq13} \\
&&  D^2_{x}f\cdot f^{*}-{\rm i}D_xg \cdot g^*+(2\alpha+1)(|f|^2 - |g|^2)=0,  \label{dnlsbilinEq14}
\end{eqnarray}
where $D$ is Hirota's bilinear differential operator. We will use these bilinear equations to derive rogue waves in the GDNLS equations (\ref{GDNLS}). It is important to notice that these bilinear equations are independent of the equation parameter $\gamma$. This means that rogue waves in the whole family of GDNLS equations (\ref{GDNLS}), for different values of $\gamma$, are given by the same $f$ and $g$ solutions, and the $\gamma$-dependence of the rogue waves only appears through the bilinear transformation (\ref{BilinearTrans}). This is a big advantage of the bilinear method for solving the GDNLS equations (\ref{GDNLS}).

Interestingly, under the same transformation (\ref{BilinearTrans}), the GDNLS equation (\ref{GDNLS}) can also be decomposed into a different bilinear system, where the first equation (\ref{dnlsbilinEq11}) is replaced by
\begin{eqnarray}
&& \left(\mathrm{i}D_{t}+D_{x}^{2}-2\mathrm{i} \alpha D_{x} \right) g\cdot f=0, \label{NewdnlsbilinEq1}
\end{eqnarray}
while the other three equations (\ref{dnlsbilinEq12})-(\ref{dnlsbilinEq14}) remain the same. This replacement is admitted because
the left side of the latter first bilinear equation (\ref{NewdnlsbilinEq1}) can be written as a linear combination of the left sides of the former bilinear equations (\ref{dnlsbilinEq11})-(\ref{dnlsbilinEq14}). Specifically, denoting the left side of each equation by its equation number, we have the identity
\[
f\times (\ref{dnlsbilinEq11})-g\times (\ref{dnlsbilinEq12})+2(\textrm{i}g_x+\alpha g) \times (\ref{dnlsbilinEq13})=f^*\times
(\ref{NewdnlsbilinEq1})-g\times (\ref{dnlsbilinEq14})+g(\textrm{i}\partial_x-1)\times (\ref{dnlsbilinEq13}).
\]
Thus, if $f$ and $g$ satisfy the former system of bilinear equations, then they would also satisfy the latter bilinear system. Although these two $(1+1)$-dimensional bilinear systems are equivalent, they have to be reduced from different higher-dimensional bilinear systems which admit different bilinear solutions. But these two different KP reductions will lead to the same rogue wave solutions, which we will show in later texts.

In this article, we will present rogue waves of the GDNLS equations (\ref{GDNLS}) through elementary Schur polynomials. These Schur polynomials $S_j(\mbox{\boldmath $x$})$ are defined by
\begin{equation*}
\sum_{j=0}^{\infty}S_j(\mbox{\boldmath $x$})\lambda^j
=\exp\left(\sum_{j=1}^{\infty}x_j\lambda^j\right),
\end{equation*}
or more explicitly,
\begin{equation*}
S_0(\mbox{\boldmath $x$})=1, \quad S_1(\mbox{\boldmath $x$})=x_1,
\quad S_2(\mbox{\boldmath $x$})=\frac{1}{2}x_1^2+x_2, \quad \cdots, \quad
S_{j}(\mbox{\boldmath $x$}) =\sum_{l_{1}+2l_{2}+\cdots+ml_{m}=j} \left( \ \prod _{j=1}^{m} \frac{x_{j}^{l_{j}}}{l_{j}!}\right),
\end{equation*}
where $\mbox{\boldmath $x$}=(x_1,x_2,\cdots)$.

\section{General rogue wave solutions}
Our general rogue wave solutions to the GDNLS equations (\ref{GDNLS}) are given by the following theorem.
\begin{quote}
\textbf{Theorem 1.} \emph{The GDNLS equations (\ref{GDNLS}) under the boundary condition  (\ref{BoundaryCond1}) admit rational rogue wave solutions}
\begin{eqnarray}
&& u_N(x,t)= e^{{\rm i}(1-\gamma-\alpha) x-{\rm i}\left[\alpha^2+2(\gamma-2)\alpha+1-\gamma\right]t}\hspace{0.05cm}\frac{(f_N^*)^{\gamma-1}g_N}{f_N^\gamma}, \label{BilinearTrans2}
\end{eqnarray}
\emph{where the positive integer $N$ represents the order of the rogue wave,}
\begin{equation*}
f_N(x,t)=\sigma_{0,0}, \quad g_N(x,t)=\sigma_{-1,1},
\end{equation*}
\begin{equation*}
\sigma_{n,k}=
\det_{
\begin{subarray}{l}
1\leq i, j \leq N
\end{subarray}
}
\left(
\begin{array}{c}
 m_{2i-1,2j-1}^{(n,k)}
\end{array}
\right),
\end{equation*}
\emph{the matrix elements in $\sigma_{n,k}$ are defined by}
\[ \label{matrixmnij}
m_{i,j}^{(n,k)}=\sum_{\nu=0}^{\min(i,j)} \frac{1}{4^{\nu}} \hspace{0.06cm} S_{i-\nu}(\textbf{\emph{x}}^{+}(n,k) +\nu \textbf{\emph{s}})  \hspace{0.06cm} S_{j-\nu}(\textbf{\emph{x}}^{-}(n,k) + \nu \textbf{\emph{s}}),
\]
\emph{vectors} $\textbf{\emph{x}}^{\pm}(n,k)=\left( x_{1}^{\pm}, x_{2}^{\pm},\cdots \right)$ \emph{are defined by}
\begin{eqnarray*}
&&x_{1}^{+}=\hspace{0.2cm}   k+ \left(n+\frac{1}{2}\right)\left(h_{1}+\frac{1}{2}\right)+\sqrt{\alpha}x+ 2\sqrt{\alpha}\left[(\alpha-1)+ {\rm i}\sqrt{\alpha}\right]t +a_{1}, \\
&&x_{1}^{-}=-k- \left(n+\frac{1}{2}\right)\left(h_{1}^*+\frac{1}{2}\right)+\sqrt{\alpha}x+ 2\sqrt{\alpha}\left[(\alpha-1)-{\rm i}\sqrt{\alpha}\right]t +a_{1}^*, \\
&&x_{r}^{+}=\hspace{0.25cm} (n+\frac{1}{2}) h_{r}+ \frac{1}{r!}\left\{\sqrt{\alpha}x+\left[2\sqrt{\alpha}(\alpha-1)+2^r {\rm i} \alpha\right]t\right\} +a_{r},  \quad r>1,  \\
&&x_{r}^{-}= -(n+\frac{1}{2})h_{r}^*+ \frac{1}{r!}\left\{\sqrt{\alpha}x+\left[2\sqrt{\alpha}(\alpha-1)-2^r {\rm i} \alpha\right]t\right\} + a_{r}^*, \quad r>1,
\end{eqnarray*}
\emph{$h_{r}(\alpha)$, $s_{r}$ are coefficients from the expansions}
\begin{eqnarray}
&& \sum_{r=1}^{\infty} h_{r}\lambda^{r}=\ln \left(\frac{{\rm i}e^{\lambda/2}+\sqrt{\alpha} e^{-\lambda/2}}{{\rm i}+\sqrt{\alpha}}\right),  \quad \sum_{r=1}^{\infty} s_{r}\lambda^{r}=\ln \left[\frac{2}{\lambda}  \tanh \left(\frac{\lambda}{2}\right)\right], \label{skrkexpcoeff2}
\end{eqnarray}
\emph{and $a_{r} \hspace{0.05cm} (r=1, 2, \dots)$ are free complex constants.}
\end{quote}

\textbf{Note 1.} The first few coefficients $h_{r}(\alpha)$ and $s_{r}$ in expansions (\ref{skrkexpcoeff2}) are
\[
h_1(\alpha)=\frac{ \textrm{i} -\sqrt{\alpha}}{2 \left(\textrm{i}+\sqrt{\alpha}\right)},
\quad h_2(\alpha)=\frac{ \textrm{i} \sqrt{\alpha }}{2 \left( \textrm{i}+\sqrt{\alpha} \right)^2},
\quad h_3(\alpha)=\frac{\sqrt{\alpha}  \left(1+i \sqrt{\alpha }\right) }{6 \left( \textrm{i}+\sqrt{\alpha} \right)^3},
\]
\[ \label{skvalues}
s_1=s_3=\dots=s_{odd}=0, \quad s_2= - \frac{1}{12}, \quad s_4=\frac{7}{1440}.
\]

Theorem 1 will be proved in Sec. \ref{sec:derivation}.

Some remarks on rogue waves in this theorem are in order. First, one can notice that the matrix-element expression in this theorem is significantly simpler than earlier such expressions for other integrable equations \cite{OhtaJY2012,OhtaJKY2012,OhtaJKY2013,OhtaJKY2014,YangYang2019Nonloc}.
Indeed, the current expression in (\ref{matrixmnij}) involves a single summation, while previous such expressions involved three summations. Second, our current parameterization of rogue waves in Theorem 1 is very different from the previous ones. In our current rogue wave solution, all internal parameters $a_1, a_2, a_3, \dots$ appear inside the $\textbf{\emph{x}}^{\pm}(n,k)$ vectors, while previous internal parameters all appeared outside such vectors as summation coefficients \cite{OhtaJY2012,OhtaJKY2012,OhtaJKY2013,OhtaJKY2014,YangYang2019Nonloc}. This different parameterization is the key reason for the simpler matrix-element expression in Theorem 1. More significantly, this parameterization facilitates the analysis of rogue waves, especially regarding the maximum peak amplitude for rogue waves of a given order. Indeed, under previous parameterizations for the NLS equation, the rogue wave with maximum peak amplitude occurs at peculiar internal parameter values \cite{OhtaJY2012}, which makes the derivation of maximum peak amplitudes at arbitrary orders intractable. However, in our current parameterization, rogue waves in Theorem 1 admit the following property.
\begin{quote}
\textbf{Theorem 2.} \emph{When $a_r=0$ for all $r\ge 1$, the rogue wave in Theorem 1 is parity-time-symmetric, i.e.,
$u_N^*(-x,-t) = u_N(x,t)$.}
\end{quote}
This property will also be proved in Sec. \ref{sec:derivation}. The significance of this property is that, this parity-time-symmetric rogue wave happens to possess the maximum peak amplitude among rogue waves of that order (see \cite{GDNLS_rogue2019}). In addition, this maximum peak amplitude is located at the center of this parity-time-symmetric rogue wave, i.e., at $x=t=0$. Thus, to derive the maximum peak amplitude of rogue waves in Theorem 1, we only need to set all its internal parameters $a_r$ as well as $(x,t)$ to zero, which is much easier. Doing so, our explicit calculations for $N=1, 2, \dots, 6$ show that
\[
|f_{N}(0,0)|_{a_r=0}=\frac{\alpha^{N(N+1)/2}}{ 2^{2 N^2} (\alpha+1)^{N(N+1)/2}},\ \  |g_{N}(0,0)|_{a_r=0}= \frac{(2N+1) \alpha^{N(N+1)/2}}{2^{2N^2}(\alpha+1)^{N(N+1)/2}},   \label{fNgN}
\]
and thus the maximum peak amplitude is
\[
|u_{N}(0,0)|_{a_r=0}=\frac{|g_{N}(0,0)|_{a_r=0}}{|f_{N}(0,0)|_{a_r=0}}=2N+1.   \label{uNmax}
\]
Remarkably, this maximum peak amplitude does not depend on the background wavenumber $\alpha$, although $|f_N|$ and $|g_N|$ in its numerator and denominator do. While these formulae (\ref{fNgN})-(\ref{uNmax}) were obtained for $N\le 6$, we believe they hold for all $N>6$ as well.

In Refs. \cite{CCL_rogue_2017,GI_rogue_2014,GDNLS_rogue2019} for the Chen-Lee-Liu equation, the Gerdjikov-Ivanov equation and the GDNLS equations (\ref{GDNLS0}), examination of some low-order rogue waves revealed that their maximum peak amplitude was $2N+1$. Our result above is more general. Interestingly, this maximum peak amplitude for rogue waves in the GDNLS equations (\ref{GDNLS}) is exactly the same as that for the NLS equation \cite{AAS2009,ACA2010,OhtaJY2012,HeNLSheight}.

Another remark on rogue waves in Theorem 1 pertains to the number of their irreducible free parameters. These rogue waves of order $N$ contain $2N-1$ complex parameters $a_1, a_2, \dots, a_{2N-1}$. But we can show that all even-indexed parameters $a_{even}$ are dummy parameters which cancel out automatically from the solution. To prove this, we first rewrite $\sigma_{n,k}$ in Theorem 1 as \cite{OhtaJY2012}
\begin{eqnarray} \label{sigmank}
&&\sigma_{n,k}=
\sum_{0\leq\nu_{1} < \nu_{2} < \cdots < \nu_{N}\leq 2N-1} \det_{1 \leq i, j\leq N} \left(
\frac{1}{2^{\nu_j}} S_{2i-1-\nu_j}(\textbf{\emph{x}}^{+}(n,k) +\nu_j \textbf{\emph{s}}) \right) \det_{1 \leq i, j\leq N}
\left(\frac{1}{2^{\nu_j}}S_{2i-1-\nu_j}(\textbf{\emph{x}}^{-}(n,k) + \nu_j \textbf{\emph{s}})\right).
\end{eqnarray}
In addition, denoting $\xi_r$ and $\eta_r$ as the real and imaginary parts of $a_r$, we can easily see that
\[ \label{Snderiv}
\partial_{\xi_r} S_{n}(\textbf{\emph{x}}^{\pm} +\nu \textbf{\emph{s}}) = S_{n-r}(\textbf{\emph{x}}^{\pm} +\nu \textbf{\emph{s}}), \quad
\partial_{\eta_r} S_{n}(\textbf{\emph{x}}^{\pm} +\nu \textbf{\emph{s}}) = \pm \mathrm{i} S_{n-r}(\textbf{\emph{x}}^{\pm} +\nu \textbf{\emph{s}}).
\]
Using these two equations, we can show that
\[ \label{dxieta2r}
\partial_{\xi_{2r}} \sigma_{n,k}=\partial_{\eta_{2r}} \sigma_{n,k}=0,
\]
which proves that rogue waves in Theorem 1 are independent of parameters $a_{even}$. Thus, we will simply set $a_{2}=a_{4}=\cdots=a_{even}=0$ throughout this article. Of the remaining parameters, we can normalize $a_{1}=0$
through a shift of $x$ and $t$. Then, the $N$-th order rogue waves in the GDNLS equation (\ref{GDNLS}) contain $N-1$ free irreducible complex parameters, $a_3, a_5, \dots, a_{2N-1}$. This number of irreducible free parameters is the same as that in rogue waves of the NLS equation \cite{OhtaJY2012}.

\section{Dynamical patterns of rogue waves}
In this section, we analyze the dynamics of rogue waves in Theorem 1 for the GDNLS equations (\ref{GDNLS}).

First of all, we notice from Eq. (\ref{BilinearTrans2}) that the amplitude profile of the rogue wave is
\[
|u_N(x,t)|=\frac{|g_N(x,t)|}{|f_N(x,t)|},
\]
which is independent of the equation parameter $\gamma$. This means that the intensity patterns of rogue waves are the same for \emph{all} GDNLS equations (\ref{GDNLS}) regardless of the $\gamma$ value. But the phase structure of rogue waves is influenced by the $\gamma$ value. Indeed, the gauge transformation (\ref{gauge}) tells us that, on top of rogue waves $v(x,t)$ of the Kaup-Newell equation, different values of $\gamma$ introduce an extra phase $\theta(x,t)=(2-\gamma)\int |v(x,t)|^2 dx$, which can be calculated directly from the bilinear solution (\ref{BilinearTrans2}). This phase induces a ``chirp" to an optical rogue wave, which was examined in detail in \cite{GDNLS_rogue2019}.

Although the rogue wave intensity pattern in the GDNLS equations (\ref{GDNLS}) is independent of $\gamma$, it does depend on the wavenumber parameter $\alpha$ of the constant background. We will focus on this $\alpha$ dependence of the rogue-wave intensity pattern next.

First, we consider fundamental rogue waves, where we set $N=1$ in Theorem 1. In addition, we normalize $a_1=0$ (see the remark in the end of the last section). Then, we get
\[ \label{Fundamplitu1}
|u_{1}(x,t; \alpha)|=\left|\frac{ \alpha (x+2 \alpha t)^2+(x-2t)^2- \textrm{i} (x+6 \alpha t) - \frac{3}{4} }{\alpha (x+2 \alpha t)^2+ (x-2t)^2+\textrm{i} (x+2 \alpha t)- 4 \textrm{i} t+ \frac{1}{4}}\right|.
\]
At three values of $\alpha=0.5, 1$ and 2, this amplitude profile is shown in Fig. 1(a,b,c) respectively. It is seen that $\alpha$ strongly affects the orientation and duration of the rogue wave. Specifically, as the $\alpha$ value increases, the orientation angle also increases, but the duration of the rogue wave decreases. However, the peak amplitudes of these rogue waves for different $\alpha$ values are all equal to 3, which are attained at the center $x=t=0$, i.e., $|u_{1}(0,0;\alpha)|= 3$. Physically, the longer duration of rogue waves at smaller $\alpha$ values can be understood, because in this case, the growth rates of baseband modulation instability can be shown to be smaller, which causes the rogue wave to take longer time to rise from the unit-amplitude background to its peak amplitude of 3. The dependence of the orientation angle on $\alpha$ can also be heuristically understood. It is known that the phase gradient of a pulse generally causes the pulse to move at a velocity which is proportional to this phase gradient. In the present case, the phase gradient of the rogue wave can be estimated from Eq. (\ref{BilinearTrans2}) as the wavenumber $1-\gamma-\alpha$. Then, for a fixed equation parameter $\gamma$, larger $\alpha$ causes the velocity to be smaller or negative, leading to a larger orientation angle. To put these results in perspective, we note that for the NLS equation, since the constant-background wavenumber of its rogue waves can be normalized by a Galilean transformation \cite{OhtaJY2012}, the background wavenumber only affects the orientation, but not duration, of its rogue waves.

\begin{figure}[htb]
   \begin{center}
   \vspace{-2.5cm}
   \includegraphics[scale=0.366, bb=0 0 385 567]{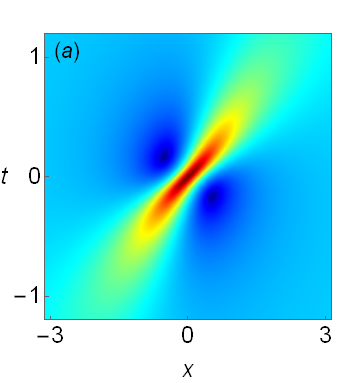}  \hspace{0.55cm}
   \includegraphics[scale=0.355, bb=0 0 385 567]{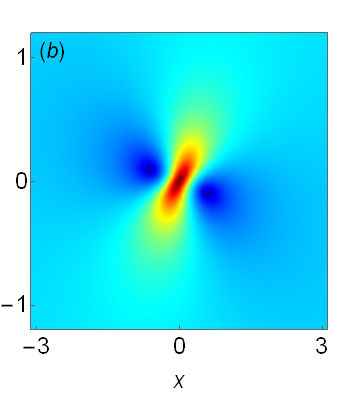}  \hspace{0.5cm}
   \includegraphics[scale=0.355, bb=0 0 385 567]{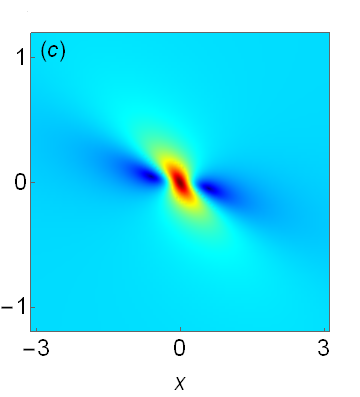}
   \caption{Amplitude profiles (\ref{Fundamplitu1}) of first-order rogue waves. (a) $\alpha=0.5$; (b) $\alpha=1$; (c) $\alpha=2$. }
   \end{center}
\end{figure}
It is interesting to notice that, in the limit of $\alpha\to 0$,
\[ \label{AlgebricSoliton1}
|u_{1}(x,t; \alpha)|\to \left|\frac{ (x-2t)^2- \textrm{i} x- \frac{3}{4} }{ (x-2t)^2+\textrm{i} x- 4 \textrm{i} t+ \frac{1}{4}}\right|,
\]
which becomes a quadratic algebraic soliton instead of a rogue wave. This comes about because when $\alpha=0$, baseband modulation instability disappears in the GDNLS equation (\ref{GDNLS}), and thus rogue waves no longer exist.

Now we consider second-order rogue waves, where we set $N=2$ and $a_{1}=0$ in Theorem 1. These solutions contain one complex free parameter $a_{3}$. When $a_3=0$, the resulting rogue wave is parity-time-symmetric, and it reaches peak amplitude 5 at the center $x=t=0$ for all $\alpha$ values, i.e., $|u_{2}(0,0; \alpha)|=5$. This peak amplitude 5 is the maximum peak amplitude for all rogue waves of second order, and thus this parity-time-symmetric rogue wave was called the super rogue wave in \cite{GDNLS_rogue2019}. The amplitude profile of this super rogue wave depends on the wavenumber parameter $\alpha$ though. At three $\alpha$ values of 0.5, 1 and 2, these super rogue waves are displayed in Fig. 2. Again, $\alpha$ strongly affects the orientation and duration of these rogue waves.

\begin{figure}[htb]
   \begin{center}
   \vspace{-2.25cm}
   \includegraphics[scale=0.350, bb=0 0 385 567]{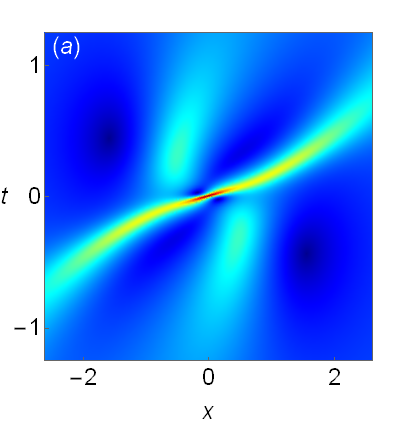} \hspace{0.6cm}
   \includegraphics[scale=0.355, bb=0 0 385 567]{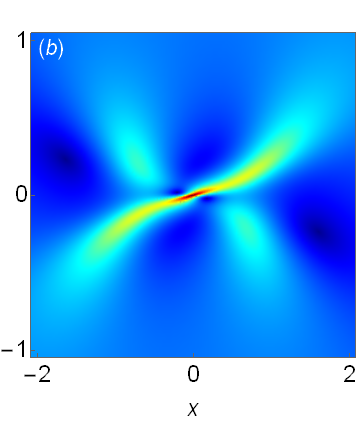}\hspace{0.5cm}
   \includegraphics[scale=0.355, bb=0 0 385 567]{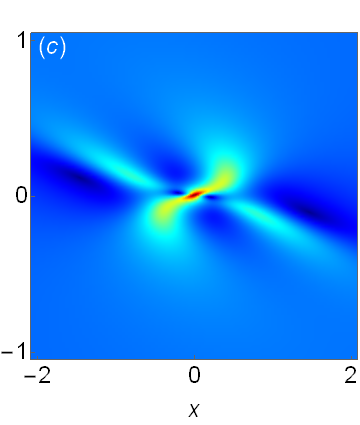}
   \caption{Amplitude profiles $|u_2(x,t)|$ of second-order super rogue waves (with $a_3=0$). (a) $\alpha=0.5$; (b) $\alpha=1$; (c) $\alpha=2$.}
   \end{center}
\end{figure}

When $a_3\ne 0$, the second-order rogue waves generally will split into three separate first-order rogue waves, as has been reported in \cite{KN_rogue_2013, CCL_rogue_2017, GI_rogue_2014, GDNLS_rogue2019}. This phenomenon is similar to second-order rogue waves of the NLS equation \cite{AAS2009,DGKM2010,ACA2010,KAAN2011,GLML2012,OhtaJY2012,DPMVB2013}. The orientations and durations of these three separate first-order rogue waves are determined by the wavenumber parameter $\alpha$.

Having clarified the effect of wavenumber parameter $\alpha$ on rogue waves, at third order, we will fix $\alpha=1$ and explore new rogue wave patterns. For this purpose, we set $N=3$ and $a_{1}=0$, and the remaining free complex parameters are $a_{3}$ and $a_{5}$. When $a_3=a_5=0$, we get a super rogue wave with peak amplitude 7 (see also \cite{CCL_rogue_2017, GI_rogue_2014, GDNLS_rogue2019}). At other $a_3$ and $a_5$ values, the third-order rogue wave generally splits into 6 separate first-order rogue waves in various configurations. Two such solutions are displayed in Fig. 3. The left panel shows a pentagon pattern, which has been seen before \cite{CCL_rogue_2017,GI_rogue_2014}. But the right panel shows a mix of a first-order rogue wave and a cluster of five first-order rogue waves in square configuration, which is novel to our knowledge. Our results suggest that when a third-order rogue wave splits into 6 separate first-order rogue waves, these 6 first-order rogue waves can appear in arbitrary configurations in the $(x,t)$ plane. The same should hold for higher-order rogue waves too.

\begin{figure}[htb]
   \begin{center}
   \vspace{-2.25cm}
   \includegraphics[scale=0.352, bb=0 0 385 567]{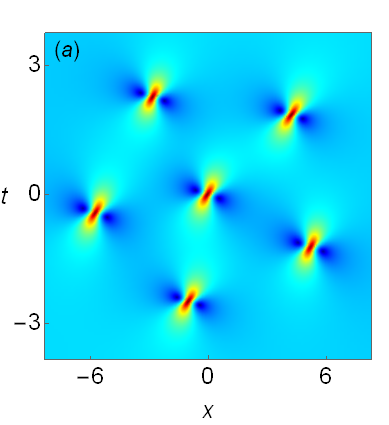}    \hspace{1.25cm}
   \includegraphics[scale=0.355, bb=0 0 385 567]{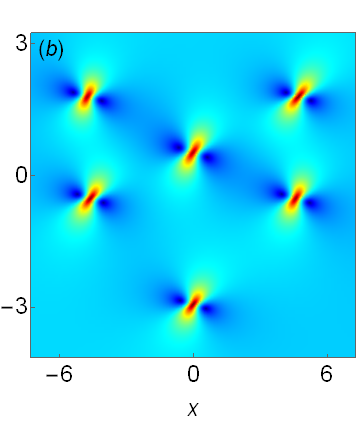}
   \caption{Third-order rogue waves with $\alpha=1$. Left: a pentagon pattern, where $a_{3}=0$ and $a_{5}=80+80\textrm{i}$. Right: a mixed pattern, where $a_{3}=10 \textrm{i}$ and $a_{5}=100 \textrm{i}$.}
   \end{center}
\end{figure}

\section{Derivation of rogue waves from the first bilinear system} \label{sec:derivation}
As we have mentioned in Sec. \ref{sec:pre}, the GDNLS equation (\ref{GDNLS}) can be decomposed into two different bilinear systems. In this section, we will derive rogue waves in Theorem 1 from the first bilinear system (\ref{dnlsbilinEq11})-(\ref{dnlsbilinEq14}). The basic idea of this derivation is similar to that in \cite{OhtaJY2012} for the NLS equation. The main improvement is that we will choose differential operators in the bilinear solutions in a different way, which leads to a more convenient parameterization and simpler expression for rogue waves.

\subsection{Gram determinant solutions for a higher-dimensional bilinear system}
First, we need to derive algebraic solutions to a higher-dimensional bilinear system, which can reduce to the original lower-dimensional bilinear system (\ref{dnlsbilinEq11})-(\ref{dnlsbilinEq14}) under certain reductions.

From Lemma 2 of Ref.~\cite{JCChen2018D-LS}, section 3.2 of Ref.~\cite{BFFengOhta2016CSP} and our own calculations, we learn that if functions $m_{i,j}^{(n,k)}$, $\varphi_{i}^{(n,k)}$ and $\psi_{j}^{(n,k)}$ of variables ($x_{-1}$, $x_{1}$, $x_{2}$) satisfy the following differential and difference relations,
\begin{eqnarray}\label{001}
\begin{array}{ll}
\partial_{x_{1}}m_{i,j}^{(n,k)}=\varphi_{i}^{(n,k)}\psi_{j}^{(n,k)},    \\
\partial_{x_{1}}\varphi_{i}^{(n,k)}=\varphi_{i}^{(n+1,k)},\  \partial_{x_{1}}\psi_{j}^{(n,k)}=-\psi_{j}^{(n-1,k)},  \\
\partial_{x_{1}}\varphi_{i}^{(n,k)}=c \varphi_{i}^{(n,k)}+ \varphi_{i}^{(n,k+1)},\  \partial_{x_{1}}\psi_{j}^{(n,k)}=-c\psi_{j}^{(n,k)}-\psi_{j}^{(n,k-1)},   \\
\partial_{x_{2}}\varphi_{i}^{(n,k)}=\partial_{x_{1}}^2 \varphi_{i}^{(n,k)},\  \partial_{x_{2}}\psi_{j}^{(n,k)}=- \partial_{x_{1}}^2 \psi_{j}^{(n,k)},   \\
\partial_{x_{-1}}\varphi_{i}^{(n,k)}=\varphi_{i}^{(n,k-1)},\  \partial_{x_{-1}}\psi_{j}^{(n,k)}=-\psi_{j}^{(n,k+1)},
\end{array}
\end{eqnarray}
where $c$ is an arbitrary complex constant, then they would also satisfy the following relations,
\begin{eqnarray}
\begin{array}{ll}
\partial_{x_{2}}m_{i,j}^{(n,k)}=\varphi_{i}^{(n+1,k)}\psi_{j}^{(n,k)}+\varphi_{i}^{(n,k)}\psi_{j}^{(n-1,k)},    \\
\partial_{x_{2}}m_{i,j}^{(n,k)}=\varphi_{i}^{(n,k+1)}\psi_{j}^{(n,k)}+\varphi_{i}^{(n,k)}\psi_{j}^{(n,k-1)}+2c\varphi_{i}^{(n,k)}\psi_{j}^{(n,k)}, \\
\partial_{x_{-1}}m_{i,j}^{(n,k)}= - \varphi_{i}^{(n,k-1)}\psi_{j}^{(n,k+1)},    \\
m_{i,j}^{(n+1, k)}=m_{i,j}^{(n,k)}+\varphi_{i}^{(n,k)}\psi_{j}^{(n+1, k)},    \\
m_{i,j}^{(n, k+1)}=m_{i,j}^{(n,k)}+\varphi_{i}^{(n,k)}\psi_{j}^{(n, k+1)}.
\end{array}
\end{eqnarray}
Using these relations, one can show that the determinant
\[\label{Mijdeterminants}
\tau_{n,k}=\det_{1\leq i,j \leq N} \left(m_{i,j}^{(n,k)}\right)
\]
would satisfy the following bilinear equations in the extended KP hierarchy
\begin{eqnarray}
&& \left(D_{x_{2}}-D_{x_{1}}^{2}-2 c D_{x_{1}}\right) \tau_{n-1, k+1} \cdot \tau_{n-1, k}=0,\label{KPHBilineEq1} \\
&& \left(D_{x_{2}}-D_{x_{1}}^{2}\right) \tau_{n, k} \cdot \tau_{n-1, k}=0, \label{KPHBilineEq2}\\
&& \left(c D_{x_{-1}}-1\right) \tau_{n, k} \cdot \tau_{n-1, k}+\tau_{n-1, k+1} \tau_{n, k-1}=0,\label{KPHBilineEq3}\\
&& (c D_{x_{1}}D_{x_{-1}} - D_{x_{1}} -2c ) \tau_{n, k} \cdot \tau_{n-1, k}+ (D_{x_{1}}+2c) \tau_{n-1, k+1} \cdot \tau_{n, k-1}=0.\label{KPHBilineEq4}
\end{eqnarray}

Now, we introduce functions $m^{(n,k)}$, $\varphi^{(n,k)}$ and $\psi^{(n,k)}$ as
\begin{eqnarray}
&& m^{(n, k)}= \frac{\mathrm{i} p}{p + q}\left(-\frac{p}{q}\right)^{n}\left(-\frac{p-c}{q+c}\right)^{k} \mathrm{e}^{\xi + \eta }, \\
&& \varphi^{(n,k)}=( \mathrm{i}p ) p^{n} (p-c)^k e^{\xi}, \\
&&  \psi^{(n,k)}=(-q)^{-n}\left[-(q+c)\right]^{-k} e^{\eta},
\end{eqnarray}
where
\begin{eqnarray}
&& \xi =\frac{1}{p-c} x_{-1}+p x_{1}+p^{2} x_{2}+\xi_{0} , \label{defxi} \\
&& \eta=\frac{1}{q+c} x_{-1}+q x_{1}-q^{2} x_{2}+ \eta_{0},  \label{defeta}
\end{eqnarray}
and $p, q, \xi_{0}$ and $\eta_{0}$ are arbitrary complex constants. It is easy to see that these functions satisfy the differential and difference relations (\ref{001}) with indices $i$ and $j$ ignored. Then, by defining
\[
m_{ij}^{(n,k)}=\mathcal{A}_i \mathcal{B}_{j} m^{(n,k)}, \quad
\varphi_i^{(n,k)}=\mathcal{A}_i\varphi^{(n,k)}, \quad
\psi_j^{(n,k)}=\mathcal{B}_{j}\psi^{(n,k)},    \label{mijn}
\]
where $\mathcal{A}_{i}$ and $\mathcal{B}_{j}$ are differential operators with respect to $p$ and $q$ respectively as
\begin{eqnarray}\label{New003b}
\mathcal{A}_{i}=\frac{1}{ i !}\left[(p-c)\partial_{p}\right]^{i}, \quad
\mathcal{B}_{j}=\frac{1}{ j !}\left[(q+c)\partial_{q}\right]^{j},
\end{eqnarray}
these functions would also satisfy the differential and difference relations (\ref{001}) since operators $\mathcal{A}_{i}$ and $\mathcal{B}_{j}$ commute with differentials $\partial_{x_k}$. Consequently, for an arbitrary sequence of indices $(i_1,i_2,\cdots,i_N; j_1,j_2,\cdots,j_N)$, the determinant
\[ \label{tildetaun}
\tau_{n,k}=\det_{1\le\nu,\mu\le N}\left( m_{i_\nu,j_\mu}^{(n,k)}\right)
\]
satisfies the higher-dimensional bilinear system (\ref{KPHBilineEq1})-(\ref{KPHBilineEq4}).

It is important to notice that the differential operators $\mathcal{A}_{i}$ and $\mathcal{B}_{j}$ defined here are simpler than the ones in previous bilinear derivations of rogue waves \cite{OhtaJY2012,OhtaJKY2012,OhtaJKY2013,OhtaJKY2014,JCChen2018LS,XiaoeYong2018,YangYang2019Nonloc}. Indeed, the current differential operators are single terms, while previous ones were defined as summations. The reason for the previous summation definitions was to introduce internal free parameters in rogue waves. In our current approach, we will introduce free constants through $\xi_{0}$ and $\eta_{0}$ in Eqs. (\ref{defxi})-(\ref{defeta}), which will be done later in this section.

Next, we will reduce the higher-dimensional bilinear system (\ref{KPHBilineEq1})-(\ref{KPHBilineEq4}) to the original bilinear system (\ref{dnlsbilinEq11})-(\ref{dnlsbilinEq14}), so that the higher-dimensional solutions (\ref{tildetaun}) become rogue wave solutions to the GDNLS equations (\ref{GDNLS}). In this reduction, we will need to set
\[ \label{akappa}
c=-\textrm{i}\alpha,
\]
where $c$ is the parameter in the higher-dimensional system (\ref{KPHBilineEq1})-(\ref{KPHBilineEq4}), and $\alpha$ is the wavenumber parameter in the original bilinear system (\ref{dnlsbilinEq11})-(\ref{dnlsbilinEq14}).

\subsection{Dimensional reduction} \label{sec:dimred}
First, we reduce the higher-dimensional bilinear system (\ref{KPHBilineEq1})-(\ref{KPHBilineEq4}) to a lower-dimensional one, a process called dimension reduction. This reduction will restrict the indices in the determinant (\ref{tildetaun}), and select the $(p,q)$ values in its matrix element $m_{i_\nu,j_\mu}^{(n,k)}$.

The dimension reduction condition we impose is
\[ \label{dimenredcdnls}
\left(\partial_{x_{1}}+ \textrm{i}c \partial_{x_{-1}} \right) \tau_{n,k}= C \tau_{n,k},
\]
where $C$ is some constant. Denoting $\hat{p}\equiv p-c$ and $\hat{q}\equiv q+c$, then $\mathcal{A}_{i}$ and $\mathcal{B}_{j}$ in Eq. (\ref{New003b}) can be rewritten as
\[ \label{AiBipqhat}
\mathcal{A}_{i}=\frac{1}{i!}\left(\hat{p}\partial_{\hat{p}}\right)^{i},\ \
\mathcal{B}_{j}=\frac{1}{j!}\left(\hat{q}\partial_{\hat{q}}\right)^{j}.
\]
In addition,
\begin{equation*}
\left(\partial_{x_{1}}+\textrm{i}c \partial_{x_{-1}} \right) m_{i,j}^{(n,k)}= \mathcal{A}_{i} \mathcal{B}_{j}\left(\partial_{x_{1}}+\textrm{i}c \partial_{x_{-1}} \right)m^{(n,k)} = \mathcal{A}_{i} \mathcal{B}_{j}\left[ \hat{p}+\frac{\textrm{i}c}{\hat{p}} + \hat{q}+\frac{\textrm{i}c}{\hat{q}} \right] m^{(n,k)}.
\end{equation*}
Using the Leibnitz rule exactly as in Ref. \cite{OhtaJY2012}, the above equation reduces to
\begin{eqnarray*}
\left(\partial_{x_{1}}+\textrm{i}c \partial_{x_{-1}} \right) m_{i,j}^{(n,k)}=  \sum_{\mu=0}^i \frac{1}{\mu !} \left(\hat{p} + (-1)^\mu \frac{\textrm{i}c}{\hat{p}} \right) m_{i-\mu,j}^{(n,k)}+\sum_{l=0}^j \frac{1}{l!}
\left( \hat{q}+ (-1)^l\frac{\textrm{i}c}{\hat{q}}  \right) m_{i,j-l}^{(n,k)}.
\end{eqnarray*}
Recalling $c=-\textrm{i}\alpha$ from (\ref{akappa}), we see that when we set $p=p_0$ and $q=q_0$, where
\[ \label{p0q0}
p_0=\sqrt{\alpha}-\textrm{i}\alpha, \quad q_0=\sqrt{\alpha}+\textrm{i}\alpha,
\]
the above equation would further simplify to
\[  \label{contigurelati}
\left(\partial_{x_{1}}+\textrm{i}c \partial_{x_{-1}} \right) \left. m_{i,j}^{(n,k)} \right|_{p=p_{0}, \ q=q_{0}}=2 \sqrt{\alpha} \sum^i_{\begin{subarray}{c} \mu=0,\\ \mu: even \end{subarray}}\frac{1}{\mu !} \left.m_{i-\mu,j}^{(n,k)}\right|_{p=p_{0}, \ q=q_{0}}+ 2 \sqrt{\alpha} \sum_{\begin{subarray}{c} l=0,\\ l: even \end{subarray}}^{j}\frac{1}{l!} \left. m_{i,j-l}^{(n,k)}\right|_{p=p_{0}, \ q=q_{0}}.
\]
Now, we restrict the general determinant (\ref{tildetaun}) to
\[ \label{tausol}
\tau_{n,k} = \det_{1\leq i, j\leq N}\left(\left. m_{2i-1,2j-1}^{(n,k)} \right|_{p=p_{0}, \ q=q_{0}} \right).
\]
Then, using the contiguity relation (\ref{contigurelati}) as in Ref. \cite{OhtaJY2012}, we get
\begin{eqnarray*}
\left(\partial_{x_{1}}+\textrm{i}c \partial_{x_{-1}} \right) \tau_{n,k} = 4 \sqrt{\alpha} \hspace{0.02cm} N \hspace{0.04cm}  \tau_{n,k},
\end{eqnarray*}
which shows that the $\tau_{n,k}$ function (\ref{tausol}) satisfies the dimension reduction condition (\ref{dimenredcdnls}).

When this dimension reduction equation is used to eliminate $x_{-1}$ from the higher-dimensional bilinear system (\ref{KPHBilineEq1})-(\ref{KPHBilineEq4}), and in view of the parameter connection (\ref{akappa}), we get
\begin{eqnarray}
&& \left(D_{x_{2}}-D_{x_{1}}^{2}+2 {\rm i}\alpha D_{x_{1}}\right) \tau_{n-1, k+1} \cdot \tau_{n-1, k}=0,\label{KPHBilineEq1b} \\
&& \left(D_{x_{2}}-D_{x_{1}}^{2}\right) \tau_{n, k} \cdot \tau_{n-1, k}=0, \label{KPHBilineEq2b}\\
&& \left({\rm i} D_{x_{1}} -1 \right) \tau_{n, k} \cdot \tau_{n-1, k}+\tau_{n-1, k+1} \tau_{n, k-1}=0, \label{KPHBilineEq3b} \\
&& (D^2_{x_{1}} + {\rm i} D_{x_{1}} +2 \alpha) \tau_{n, k} \cdot \tau_{n-1, k}- ({\rm i} D_{x_{1}}+2 \alpha) \tau_{n-1, k+1}  \cdot \tau_{n, k-1}=0. \label{KPHBilineEq4b}
\end{eqnarray}
In addition, using Eq. (\ref{KPHBilineEq3b}), we can replace the last bilinear equation (\ref{KPHBilineEq4b}) by
\begin{eqnarray}
D^2_{x_{1}}  \tau_{n, k} \cdot \tau_{n-1, k} - {\rm i}  D_{x_{1}} \tau_{n-1, k+1}  \cdot \tau_{n, k-1}  +(2\alpha+1) (\tau_{n, k} \cdot \tau_{n-1, k}-\tau_{n-1, k+1}  \cdot \tau_{n, k-1})=0.   \label{KPHBilineEq4bb}
\end{eqnarray}
In these reduced bilinear equations, the $x_{-1}$ derivative disappears.

To further reduce the bilinear system (\ref{KPHBilineEq1b})-(\ref{KPHBilineEq3b}) and (\ref{KPHBilineEq4bb}) to the original system (\ref{dnlsbilinEq11})-(\ref{dnlsbilinEq14}), we set
\[
x_1=x-2t, \quad x_2={\rm i}t.
\]
Under this variable relation, we have
\[
\partial_{x_{1}}=\partial_{x}, \quad  \partial_{x_{2}}= -\textrm{i} \partial_{t} -2  \textrm{i} \partial_{x}.
\]
Inserting these equations into the bilinear system (\ref{KPHBilineEq1b})-(\ref{KPHBilineEq3b}) and (\ref{KPHBilineEq4bb}), and setting $n=k=0$, we get
\begin{eqnarray}
&&\left(\mathrm{i}D_{t}+D_{x}^{2}+2\mathrm{i}(1-\alpha)D_{x} \right) g\cdot \bar{f}=0, \label{KPHBilineEq1c}\\
&&\left({\rm i} D_t + D^2_x +2{\rm i}  D_x   \right) f\cdot \bar{f}=0, \label{KPHBilineEq2c}\\
&&({\rm i}D_{x}-1) f\cdot \bar{f}+g \bar{g}=0,   \label{KPHBilineEq3c} \\
&&D_{x}^2f\cdot \bar{f}-{\rm i}D_xg \cdot \bar{g} +(2\alpha+1)(|f|^2-|g|^2)=0, \label{KPHBilineEq4c}
\end{eqnarray}
where $f, \bar{f}, g$ and $\bar{g}$ are defined as
\[ \label{fgdef1}
f=\tau_{0,0}, \quad \bar{f}=\tau_{-1,0}, \quad  g=\tau_{-1,1}, \quad \bar{g}=\tau_{0,-1}.
\]

\subsection{Complex conjugacy conditions}
Next, we need to impose complex conjugacy conditions $\bar{f}=f^*$ and $\bar{g}=g^*$, i.e.,
\[ \label{ComplConjuTaunf}
\tau_{-1,0}=\tau_{0,0}^*, \quad  \tau_{0,-1}=\tau_{-1,1}^*,
\]
so that the bilinear system (\ref{KPHBilineEq1c})-(\ref{KPHBilineEq4c}) would reduce to the original bilinear system (\ref{dnlsbilinEq11})-(\ref{dnlsbilinEq14}). These conditions can be satisfied by imposing the parameter constraint $\xi_{0}=\eta_{0}^*$.
Indeed, under this constraint, since $x_1=x-2t$ is real, $x_2=\textrm{i}t$, $c=-{\rm i} \alpha $ are pure imaginary, and $q_{0}=p_{0}^*$, we can easily show that
\[ \label{mijsym}
\left. \left[m_{i,j}^{(n,k)}\right]^* \right|_{p=p_{0}, \ q=q_{0}}=  \left. m_{j,i}^{(-n-1, -k)}\right|_{p=p_{0}, \ q=q_{0}}.
\]
Thus, $\tau_{n,k}^*=\tau_{-n-1,-k}$, i.e., the complex conjugacy conditions (\ref{ComplConjuTaunf}) hold.

\subsection{Rogue wave solutions in differential operator form}
Finally, we need to introduce free parameters into rogue waves. Unlike all previous bilinear approaches \cite{OhtaJY2012,OhtaJKY2012,OhtaJKY2013,OhtaJKY2014,JCChen2018LS,XiaoeYong2018,YangYang2019Nonloc}, we will introduce free parameters through the arbitrary constant $\xi_0$ in Eq. (\ref{defxi}). Specifically, we choose $\xi_0$ as
\[ \label{defxi0}
\xi_0=\sum _{r=1}^\infty \hat{a}_{r} \ln^{r}\left(\frac{p-c}
{p_0-c}\right)=\sum _{r=1}^\infty \hat{a}_{r} \ln^{r}\left(\frac{p+{\rm i}\alpha}
{\sqrt{\alpha}}\right),
\]
where $\hat{a}_r$ are free complex constants. We can show that rogue waves with this new parameterization can be converted to those with the old parameterization through nontrivial parameter connections. But the new parameterization will drastically simplify rogue wave expressions.

Putting all the above results together and setting $x_{-1}=0$, rational solutions to the GDNLS equations (\ref{GDNLS}) are given by the following theorem.
\begin{quote}
\textbf{Theorem 3} \emph{The GDNLS equations (\ref{GDNLS}) admit rational solutions}
\begin{eqnarray}
&& u_N(x,t)= e^{{\rm i}(1-\gamma-\alpha) x-{\rm i}\left[\alpha^2+2(\gamma-2)\alpha+1-\gamma\right]t}\hspace{0.05cm}\frac{(f_N^*)^{\gamma-1}g_N}{f_N^\gamma},
\end{eqnarray}
\emph{where}
\[
f_N(x,t)=\tau_{0,0}, \quad g_N(x,t)=\tau_{-1,1},
\]
\[
\tau_{n,k}=
\det_{
\begin{subarray}{l}
1\leq i, j \leq N
\end{subarray}
}
\left(
\begin{array}{c}
m_{2i-1,2j-1}^{(n,k)}
\end{array}
\right),
\]
\emph{the matrix elements in $\tau_{n,k}$ are defined by}
\begin{eqnarray} \label{mij-diff}
&& m_{i,j}^{(n,k)}=
  \frac{\left[(p + {\rm i} \alpha )\partial_{p}\right]^{i}}{ i !}\frac{\left[(q-{\rm i} \alpha )\partial_{q}\right]^{j}}{ j !} \left. \left[\frac{\textrm{i} p}{p+q} \left(- \frac{p}{q} \right)^n \left(- \frac{p+{\rm i} \alpha }{q-{\rm i} \alpha } \right)^k
e^{\Theta(x,t)}\right]\ \right|_{p=p_{0}, \ q=q_{0}},
\end{eqnarray}
\emph{with}
\begin{eqnarray}
&& \Theta(x,t)= (p+q) (x-2t)+ (p^2-q^2)  \textrm{i} t +\sum _{r=1}^\infty \hat{a}_{r} \ln^{r}\left(\frac{p+{\rm i}\alpha}
{\sqrt{\alpha}}\right)+ \sum _{r=1}^\infty \hat{a}^*_{r} \ln^{r}\left(\frac{q-{\rm i} \alpha}{\sqrt{\alpha}}\right),
\end{eqnarray}
\emph{$p_0, q_0$ are given in Eq. (\ref{p0q0}), $\alpha>0$, and $\hat{a}_{r} \hspace{0.05cm} (r=1, 2, \dots)$ are free complex constants.}
\end{quote}

\subsection{Explicit rogue wave solutions through Schur polynomials} \label{SchurPolynoExp}
The above rational solutions in Theorem 3 involve differential operators, which make them less explicit. More seriously, such forms make analysis of those solutions difficult. For instance, under such forms, it is difficult to prove that they satisfy the boundary conditions (\ref{BoundaryCond1}). In addition, it is difficult to determine the maximum peak amplitudes for rogue waves of each order. Thus, in this subsection, we derive a more explicit form for these solutions, which is the one given in Theorem 1 earlier in the paper.

The technique we use is similar to that in Ref. \cite{OhtaJY2012}. The differential operators in (\ref{mij-diff}) can be rewritten as (\ref{AiBipqhat}), where $\hat{p}=p+\textrm{i}\alpha$ and $\hat{q}=q-\textrm{i}\alpha$, and the $m^{(n,k)}$ term following the differential operators in (\ref{mij-diff}) can be rewritten as
\begin{eqnarray*}
&& m^{(n,k)}=\frac{\textrm{i} (\hat{p}- \textrm{i} \alpha)}{\hat{p}+\hat{q}} \left(- \frac{ \hat{p}- \textrm{i} \alpha}{ \hat{q}+\textrm{i} \alpha} \right)^n \left(-\frac{\hat{p}}{\hat{q}} \right)^k \times   \nonumber \\
&&\hspace{1.1cm} \exp\left\{\left(\hat{p}+ \hat{q}\right) (x-2t)+ \left[\hat{p}^2-\hat{q}^2 - 2 \textrm{i} \alpha (\hat{p}+\hat{q})\right] \textrm{i}t +
\sum _{r =1}^\infty \hat{a}_{r } \ln^{r }\left(\frac{\hat{p}}{\hat{p}_{0}}\right)+\sum _{r =1}^\infty \hat{a}^*_{r} \ln^{r}\left(\frac{\hat{q}}{\hat{q}_{0}}\right)\right\},
\end{eqnarray*}
where $\hat{p}_{0}=p_0+\textrm{i}\alpha$ and $\hat{q}_0=q_0-\textrm{i}\alpha$, i.e., $\hat{p}_{0}=\hat{q}_0=\sqrt{\alpha}$.
Then, introducing the generator $\mathcal{G}$ of differential operators $\left(\hat{p} \partial_{\hat{p}}\right)^{i}  \left(\hat{q} \partial_{\hat{q}}\right)^{j}$ as
\[ \label{GeneratorG}
\mathcal{G}= \sum_{i=0}^\infty \sum_{j=0}^{\infty} \frac{\zeta^i}{i!} \frac{\lambda^j}{j!} \left[\hat{p} \partial_{\hat{p}} \right]^{i}  \left[ \hat{q} \partial_{\hat{q}}\right]^{j},
\]
and utilizing the formula \cite{OhtaJY2012}
\[ \label{Gdef}
\mathcal{G}F(\hat{p}, \hat{q}) = F\left(e^{\zeta}\hat{p}, e^{\lambda}\hat{q}\right),
\]
we get
\begin{eqnarray*}
&&\left. \mathcal{G} m^{(n,k)}\right|_{\hat{p}=\hat{p}_{0},\ \hat{q} =\hat{q}_{0}} =
\frac{ e^{\zeta/2}(\textrm{i}  e^{\zeta/2} +  \sqrt{\alpha}e^{-\zeta/2} )}{e^{\zeta}+e^{\lambda}}
(-1)^k e^{(k+\frac{n}{2})(\zeta-\lambda)}
\left( \frac{\textrm{i}  e^{\zeta/2} +  \sqrt{\alpha}e^{-\zeta/2 }}{- \textrm{i}  e^{\lambda/2} +  \sqrt{\alpha}e^{-\lambda/2}} \right)^n \times \\
&& \hspace{2.7cm} \exp\left\{\sqrt{\alpha} \left( e^\zeta  + e^\lambda \right) (x-2t +2\alpha t)+ \alpha \left( e^{2\zeta} - e^{2\lambda}\right) \textrm{i}t  +\sum _{r =1}^\infty \left(a_{r } \zeta^{r }+a^*_{r}\lambda^{r}\right) \right\}.
\end{eqnarray*}
Since
\begin{eqnarray*}
\left. m^{(n,k)} \right|_{\hat{p}=\hat{p}_{0},\ \hat{q} =\hat{q}_{0}} = \left(- 1 \right)^k  \frac{ (\textrm{i}+ \sqrt{\alpha} )}{2} \left(\frac{ \textrm{i}+\sqrt{\alpha}}{ - \textrm{i}+\sqrt{\alpha}} \right)^n e^{ 2 \sqrt{\alpha} \left(x-2t +2 \alpha t\right)},
\end{eqnarray*}
we have
\begin{eqnarray}
&& \frac{1}{m^{(n,k)}}\left. \mathcal{G} m^{(n,k)}\right|_{\hat{p}=\hat{p}_{0},\ \hat{q} =\hat{q}_{0}}=
\frac{2}{e^{\zeta}+e^{\lambda}} e^{\zeta/2+(k+\frac{n}{2})(\zeta-\lambda)}
\left(\frac{\textrm{i}e^{\zeta/2} +\sqrt{\alpha}e^{-\zeta/2}}{\textrm{i}+\sqrt{\alpha}}\right)^{n+1}
\left(\frac{-\textrm{i}+\sqrt{\alpha}}{-\textrm{i}  e^{\lambda/2} +  \sqrt{\alpha}e^{-\lambda/2}}\right)^n \times \nonumber  \\
&& \hspace{3.6cm} \exp\left(\sqrt{\alpha} \left( e^\zeta  + e^\lambda -2 \right) \left( x-2t +  2  \alpha t  \right) +\alpha\left( e^{2\zeta}  - e^{2\lambda} \right)\textrm{i}t +\sum _{r =1}^\infty (a_{r} \zeta^{r} +a^*_{r} \lambda^{r}) \right) \label{AlgeExpress1}.
\end{eqnarray}
Now, we need to expand the right side of the above equation into power series of $\zeta$ and $\lambda$. For this purpose, we denote
\begin{equation*}
\frac{\textrm{i}  e^{\zeta/2} +  \sqrt{\alpha}e^{-\zeta/2}}{ \textrm{i}+ \sqrt{\alpha}}=\exp\left[ \ln \left( \frac{\textrm{i}  e^{\zeta/2} +  \sqrt{\alpha}e^{-\zeta/2}}{ \textrm{i}+ \sqrt{\alpha}} \right) \right] =\exp\left(\sum_{r=1}^{\infty} h_{r}\zeta^{r}\right),
\end{equation*}
where $h_r(\alpha)$ is as defined in Eq. (\ref{skrkexpcoeff2}). The exponent in the most right-hand side of Eq. (\ref{AlgeExpress1}) can be rewritten as
\begin{equation*}
\exp \left( \sum_{r=1}^{\infty}\frac{\zeta^r}{r!} \left( \sqrt{\alpha} (x-2t+ 2  \alpha t) +2^r  \textrm{i} \alpha t \right) +  \sum_{r=1}^{\infty}\frac{\lambda^r}{r!} \left( \sqrt{\alpha} (x-2t+ 2 \alpha t) -2^r \textrm{i} \alpha t \right) +\sum _{r =1}^\infty (a_{r}\zeta^{r}+a^*_{r} \lambda^{r})  \right),
\end{equation*}
and the $2/(e^{\zeta}+e^{\lambda})$ term can be written as \cite{OhtaJY2012}
\begin{equation*}
  \frac{2}{e^{\zeta}+e^{\lambda}}=  \sum_{\nu=0}^{\infty} \left(\frac{\zeta\lambda}{4} \right)^{\nu} \exp\left(  \sum_{r=1}^{\infty} \left(  \nu s_{r}- c_{r}\right)\left(\zeta^r + \lambda^{r}\right) - \frac{\zeta}{2}-\frac{\lambda}{2} \right),
\end{equation*}
where $c_r$ are Taylor coefficients of $\lambda^r$ in the expansion of $\ln\cosh(\lambda/2)$,
and $s_r$ are given in Eq. (\ref{skrkexpcoeff2}). Combining the above results, Eq. (\ref{AlgeExpress1}) becomes
\begin{eqnarray} \label{Gmnk}
&&\frac{1}{m^{(n,k)}}\left. \mathcal{G} m^{(n,k)}\right|_{\hat{p}=\hat{p}_{0},\ \hat{q} =\hat{q}_{0}} =\sum_{\nu=0}^{\infty} \left( \frac{\zeta\lambda}{4} \right)^{\nu} \exp\left(  \sum_{r=1}^{\infty} \left(x_{r}^{+}+\nu s_{r}\right) \zeta^r + \sum_{r=1}^{\infty} \left(x_{r}^{-}+\nu s_{r}\right) \lambda^{r} \right),
\end{eqnarray}
where $x_{r}^{+}(n,k)$ and $x_{r}^{-}(n,k)$ are defined as
\begin{eqnarray*}
&&x_{1}^{+}(n,k)=\sqrt{\alpha} (x-2t+2\alpha t)+2\textrm{i}\alpha t +(n+1) h_1 +k+\frac{n}{2}-c_1+\hat{a}_1, \\
&&x_{1}^{-}(n,k)=\sqrt{\alpha} (x-2 t+2\alpha t)-2\textrm{i} \alpha t-n h^*_1-k-\frac{1}{2} (n+1)-c_1+\hat{a}^*_{1}, \\
&&x_{r}^{+}(n,k)=\frac{1}{r !}\left[\sqrt{\alpha} (x-2 t+2\alpha t)+2^{r} \textrm{i}\alpha t  \right]+(n+1)h_{r} -c_{r} +\hat{a}_{r},\\
&&x_{r}^{-}(n,k)= \frac{1}{r !}\left[\sqrt{\alpha} (x-2 t+2\alpha t)-2^{r}\textrm{i} \alpha t \right]-nh^*_{r}-c_{r}+ \hat{a}^*_{r}.
\end{eqnarray*}
We further define shifted parameters
\begin{eqnarray*}
&& a_{1}=\hat{a}_{1}-c_{1}+\frac{1}{2}h_{1}-\frac{1}{4}, \quad
a_{r}=\hat{a}_{r}-c_{r}+\frac{1}{2}h_{r}.
\end{eqnarray*}
Then the above $x_{r}^{+}$ and $x_{r}^{-}$ reduce to those in Theorem 1. Taking the coefficients of $\zeta^i \lambda^j$ on both sides of Eq. (\ref{Gmnk}), we get
\begin{equation*}
\frac{m_{i,j}^{(n,k)}}{\left. m^{(n,k)}\right|_{p=p_{0}, q=q_{0}}}
=\sum_{\nu=0}^{\min(i,j)} \frac{1}{4^\nu} S_{i-\nu}\left( \textbf{\emph{x}}^{+}+\nu \textbf{\emph{s}} \right)
S_{j-\nu}\left( \textbf{\emph{x}}^{-}+\nu \textbf{\emph{s}} \right),
\end{equation*}
where $m_{i,j}^{(n,k)}$ is the matrix element defined in Eq. (\ref{mij-diff}) of Theorem 3. Finally, we define
\begin{equation*}
\sigma_{n,k}=\frac{\tau_{n,k}}{\left(\left. m^{(n,k)}\right|_{p=p_{0}, q=q_{0}}\right)^N}.
\end{equation*}
Then the matrix element in $\sigma_{n,k}$ is as given in Theorem 1. Since the bilinear equations (\ref{dnlsbilinEq11})-(\ref{dnlsbilinEq14}) are invariant when $f$ and $g$ are divided by an arbitrary complex constant multiplying an exponential of a linear and real function in $x$ and $t$, $\sigma_{n,k}$ then is also a solution to the GDNLS equations (\ref{GDNLS}).

Regarding boundary conditions of these rational solutions, using the Schur polynomial expressions in Theorem 1 and the same technique as in Ref. \cite{OhtaJY2012}, we can show that when $x$ or $t$ approaches infinity, $f_N(x,t)$ and $g_N(x,t)$ have the same leading term, which is also real. Thus, the rational solution (\ref{BilinearTrans2}) satisfies the boundary condition (\ref{BoundaryCond1}), and is thus a rogue wave.
Theorem 1 is then proved.

\subsection{The parity-time-symmetric rogue wave}
In this subsection, we derive the parity-time-symmetric rogue wave and prove Theorem 2.

When we set all $a_{r}=0$ in Theorem 1, $x_r^+$  and $x_r^{-}$ satisfy the following relations
\begin{equation*}
\widehat{x}_r^{\pm}(x,t) = - x_r^{\mp}(x,t), \quad r \geq 1,
\end{equation*}
where we have defined $\widehat{f}(x,t)\equiv f^*(-x,-t)$ for any function $f(x,t)$. Thus,
\begin{equation*}
\widehat{\textbf{\emph{x}}}^{\pm}(n,k)+\nu\ \textbf{\emph{s}}=\textbf{y}^{\mp}(n,k)+\nu\ \textbf{\emph{s}}+\textbf{z}^{\mp}(n),
\end{equation*}
where vectors $\textbf{y}^{\pm}$ and $\textbf{z}^{\pm}$ are defined as
\begin{equation*}
\textbf{y}^{\pm}= \left(-x_{1}^{\pm}, x_{2}^{\pm}, -x_{3}^{\pm}, x_{4}^{\pm}, \cdots\right), \quad
\textbf{z}^{\pm}=\left(0, -2x_{2}^{\pm}, 0, -2x_{4}^{\pm}, 0,  \cdots\right).
\end{equation*}
Notice that
\begin{eqnarray*}
&& \sum_{j=0}^{\infty} S_{j}\left(\widehat{\textbf{\emph{x}}}^{\mp}+\nu\ \textbf{\emph{s}}\right)\lambda^{j} =\sum_{j=0}^{\infty} S_{j}\left(\textbf{y}^{\pm}+\nu\ \textbf{\emph{s}}+\textbf{z}^{\pm}\right)\lambda^{j}  \\
&& = \exp\left(\sum_{j=1}^{\infty} \left(y_j^{\pm}+\nu\ s_j+z_j^{\pm}\right)\lambda^j\right)
=\exp\left(\sum_{j=1}^{\infty} \left(y_j^{\pm}+\nu\ s_j\right)\lambda^j\right)\exp\left(\sum_{j=1}^{\infty} z_j^{\pm}\lambda^j\right) \\
&& = \sum_{j=0}^{\infty} S_{j}(\textbf{y}^{\pm}+\nu \textbf{\emph{s}})\lambda^{j} \sum_{j=0}^{\infty} S_{j}(\textbf{z}^{\pm}) \lambda^{j}=\sum_{j=0}^{\infty} \sum_{\mu_{1}+\mu_{2}=j} S_{\mu_{1}}(\textbf{y}^{\pm}+\nu \textbf{\emph{s}})S_{\mu_{2}}(\textbf{z}^{\pm}) \lambda^{j}.
\end{eqnarray*}
Since $s_1=s_3=\dots=s_{odd}=0$ in view of Eq. (\ref{skvalues}), by comparing the coefficient of $\lambda^{j}$ on the two sides of this equation and utilizing Lemmas 2 and 3 in Ref. \cite{YangYang2019Nonloc}, we get the relation
\begin{eqnarray}\label{comparpowern}
S_{j}\left(\widehat{\textbf{\emph{x}}}^{\mp}+\nu\ \textbf{\emph{s}}\right)= (-1)^{j} \sum_{\mu=0}^{\left[ j/2 \right]} S_{\mu}(\textbf{\emph{w}}^\pm) S_{j-2\mu}(\textbf{\emph{x}}^{\pm}+\nu \textbf{\emph{s}}),
\end{eqnarray}
where $\textbf{\emph{w}}^\pm=\left( -2x_{2}^{\pm}, -2x_{4}^{\pm},\cdots \right)$. Recall from Theorem 1 that
\begin{eqnarray*}
&& \sigma_{n,k}= \det_{1 \leq i, j\leq N}
\left(\sum_{\nu=0}^{\min\left(2i-1, \ 2j-1\right)}
\frac{1}{2^{\nu}} S_{2i-1-\nu}(\textbf{\emph{x}}^{+}(n,k) +\nu \textbf{\emph{s}})
\frac{1}{2^{\nu}} S_{2j-1-\nu}(\textbf{\emph{x}}^{-}(n,k) + \nu \textbf{\emph{s}})
\right),
\end{eqnarray*}
and $\widehat{\sigma}_{n,k}$ is equal to the right side of the above equation with $\textbf{\emph{x}}^{\pm}$ replaced by $\widehat{\textbf{\emph{x}}}^{\pm}$.
By rewriting these two determinants into $3N\times 3N$ determinants as in \cite{OhtaJY2012}, utilizing relations (\ref{comparpowern}) and performing simple row manipulations, we can quickly show that
$\widehat{\sigma}_{n,k} = \sigma_{n,k}$. Thus, the solution $u_N(x,t)$ in Theorem 1 with all $a_r$ being zero satisfies the parity-time symmetry $\widehat{u}_N=u_N$, i.e., $u_N^*(-x,-t)=u_N(x,t)$. Theorem 2 is then proved.

It turns out that the converse is also true, i.e., if a rogue wave $u_N(x,t)$ in Theorem 1 is parity-time-symmetric, then $a_1=a_3=\dots=a_{odd}=0$ [there is no restriction on the $a_{even}$ values because the solution is independent of them, see Eq. (\ref{dxieta2r})].
Our proof is based on calculating the derivatives of the polynomial $\sigma_{n,k}$ with respect to the real part $\xi_{2r-1}$ and imaginary part $\eta_{2r-1}$ of the parameter $a_{2r-1}$. Using Eqs. (\ref{sigmank})-(\ref{Snderiv}), we can show that each of
$\partial_{\xi_{2r-1}} \sigma_{n,k}$ and $\textrm{i}\partial_{\eta_{2r-1}} \sigma_{n,k}$ contains power terms of $(x,t)$ which are not parity-time-symmetric. Thus, if any $a_{odd}$ is non-zero, the solution $u_N(x,t)$ would not be parity-time-symmetric.

\section{Rogue waves through a different KP-reduction procedure}
As we have mentioned in Sec. \ref{sec:pre}, the GDNLS equations (\ref{GDNLS}) admit two different bilinearizations. The first bilinear system is Eqs. (\ref{dnlsbilinEq11})-(\ref{dnlsbilinEq14}), while the second bilinear system is Eqs. (\ref{dnlsbilinEq12})-(\ref{dnlsbilinEq14}) and (\ref{NewdnlsbilinEq1}), i.e.,
\begin{eqnarray}
&& \left(\mathrm{i}D_{t}+D_{x}^{2}-2\mathrm{i} \alpha D_{x} \right) g\cdot f=0, \label{NewdnlsbilinEq1b}\\
&& \left({\rm i} D_t + D^2_x +2{\rm i}  D_x \right) f\cdot f^*=0, \label{NewdnlsbilinEq2b}\\
&& \left(\mathrm{i} D_{x}-1\right) f\cdot f^{*}+ |g|^{2}=0 \label{NewdnlsbilinEq3b} \\
&& D^2_{x}f\cdot f^{*}-{\rm i}D_xg \cdot g^*+(2\alpha+1)(|f|^2 - |g|^2)=0.  \label{NewdnlsbilinEq4b}
\end{eqnarray}
Rogue waves in the GDNLS equations (\ref{GDNLS}), as given in Theorem 1, can also be derived from this second bilinear system, but the corresponding KP-reduction procedure is different. This will be shown below. This situation is analogous to multi-soliton solutions in the Sasa-Satsuma equation, which also admit two different bilinearizations and two different reduction procedures \cite{SS_twosets}

\subsection{Algebraic solutions for a higher-dimensional bilinear system}
First, we consider the following higher-dimensional bilinear equations in the extended KP hierarchy
\begin{eqnarray}
&& \left(D_{x_{2}}-D_{x_{1}}^{2}-2d D_{x_{1}}\right) \tau_{n, k, l+1} \cdot \tau_{n, k, l}=0, \label{NewKPBilineEq1}\\
&& \left(D_{x_{2}}-D_{x_{1}}^{2}\right) \tau_{n, k, l} \cdot \tau_{n-1, k,l}=0,\label{NewKPBilineEq2} \\
&& \left(c D_{x_{-1}}+1\right) \tau_{n-1, k, l} \cdot \tau_{n, k, l}=\tau_{n, k-1, l} \tau_{n-1, k+1, l},\label{NewKPBilineEq3}\\
&& ( c D_{x_{1}}D_{x_{-1}} - D_{x_{1}} - 2c ) \tau_{n, k, l} \cdot \tau_{n-1, k, l}= (D_{x_{1}}-2c) \tau_{n,k-1, l} \cdot
\tau_{n-1, k+1, l},\label{NewKPBilineEq4}
\end{eqnarray}
where $c$ and $d$ are arbitrary complex constants. The main difference between these bilinear equations and the previous ones (\ref{KPHBilineEq1})-(\ref{KPHBilineEq4}) is the introduction of the third index $l$ in the $\tau$ function, which is necessary in order to reduce the first bilinear equation (\ref{NewKPBilineEq1}) to (\ref{NewdnlsbilinEq1b}). Indeed, the previous two-index $\tau$ function (\ref{Mijdeterminants}) is unable to satisfy a higher-dimensional bilinear equation which can be reduced to (\ref{NewdnlsbilinEq1b}).

We can show that if functions $m_{i,j}^{(n,k,l)}$, $\varphi_{i}^{(n,k,l)}$ and $\psi_{j}^{(n,k,l)}$ of variables ($x_{-1}$, $x_{1}$, $x_{2}$)  satisfy the following differential and difference relations
\[ \label{DD1}
\begin{array}{ll}
\partial_{x_{1}}m_{i,j}^{(n,k,l)}=\varphi_{i}^{(n,k,l)}\psi_{j}^{(n,k,l)},  \\
\partial_{x_{1}}\varphi_{i}^{(n,k,l)}= \varphi_{i}^{(n+1,k,l)},\  \partial_{x_{1}}\psi_{j}^{(n,k,l)}=-\psi_{j}^{(n-1,k,l)}, \\
\partial_{x_{1}}\varphi_{i}^{(n,k,l)}=c \varphi_{i}^{(n,k,l)}+ \varphi_{i}^{(n,k+1,l)},\  \partial_{x_{1}}\psi_{j}^{(n,k,l)}=-c\psi_{j}^{(n,k,l)}-\psi_{j}^{(n,k-1,l)},      \\
\partial_{x_{1}}\varphi_{i}^{(n,k,l)}= d \varphi_{i}^{(n,k,l)}+ \varphi_{i}^{(n,k,l+1)},\  \partial_{x_{1}}\psi_{j}^{(n,k,l)}=-d \psi_{i}^{(n,k,l)}-\psi_{j}^{(n,k,l-1)}, \\
\partial_{x_{2}}\varphi_{i}^{(n,k,l)}=\partial_{x_{1}}^2 \varphi_{i}^{(n,k,l)},\  \partial_{x_{2}}\psi_{j}^{(n,k,l)}=- \partial_{x_{1}}^2 \psi_{j}^{(n,k,l)}, \\
\partial_{x_{-1}}\varphi_{i}^{(n,k,l)}=\varphi_{i}^{(n,k-1,l)},\  \partial_{x_{-1}}\psi_{j}^{(n,k,l)}=-\psi_{j}^{(n,k+1,l)},
\end{array}
\]
then the determinant
\[\label{TaufuncnNKL}
\tau_{n, k, l}=\det_{1\leq i,j \leq N} \left(m_{i,j}^{(n,k,l)}\right)
\]
would satisfy the new higher-dimensional bilinear system (\ref{NewKPBilineEq1})-(\ref{NewKPBilineEq4}).

Now, we introduce the function $m^{(n,k,l)} $ as
\begin{eqnarray*}
&& m^{(n,k,l)} = \frac{\mathrm{i} p }{p + q}\left(-\frac{p}{q}\right)^{n} \left(-\frac{p-c}{q+c}\right)^{k}
\left(-\frac{p-d}{q+d}\right)^{l} \mathrm{e}^{\xi + \eta },
\end{eqnarray*}
where
\begin{eqnarray*}
&& \xi =\frac{1}{p-c} x_{-1}+p x_{1}+p^{2} x_{2}+\xi_{0} ,\\
&& \eta=\frac{1}{q+c} x_{-1}+q x_{1}-q^{2} x_{2}+ \eta_{0},
\end{eqnarray*}
and $\xi_{0}$ and $\eta_{0}$ are arbitrary complex constants. Then, by defining
\[
m_{i,j}^{(n,k,l)}=\mathcal{A}_i \mathcal{B}_{j} m^{(n,k,l)}, \label{mijnklfunc}
\]
where $\mathcal{A}_{i}$ and $\mathcal{B}_{j}$ are differential operators as defined in Eq. (\ref{New003b}), then this $m_{i,j}^{(n,k,l)}$, together with appropriately chosen $\varphi_{i}^{(n,k,l)}$ and $\psi_{j}^{(n,k,l)}$, satisfies those differential-difference equations (\ref{DD1}), and thus the determinant (\ref{TaufuncnNKL}) satisfies the bilinear system (\ref{NewKPBilineEq1})-(\ref{NewKPBilineEq4}) for arbitrary sequences of indices $(i_1,i_2,\cdots,i_N; j_1,j_2,\cdots,j_N)$.

To reduce the higher-dimensional bilinear system (\ref{NewKPBilineEq1})-(\ref{NewKPBilineEq4}) to (\ref{NewdnlsbilinEq1b})-(\ref{NewdnlsbilinEq4b}), we will set
\[ \label{cdvalues}
c=-{\rm i} \alpha, \quad d=-{\rm i} (1+\alpha).
\]

\subsection{Dimension reduction}
Our dimension reduction is the same as before, i.e.,
\[ \label{dimenredcdnls2}
\left[\partial_{x_{1}}+ \textrm{i} c \partial_{x_{-1}} \right] \tau_{n,k,l}= C  \tau_{n,k,l},
\]
where $C$ is a certain constant. The same calculations as in Sec. \ref{sec:dimred} show that the determinant
\[ \label{tausol2}
\tau_{n,k,l} = \det_{1\leq i, j\leq N}\left(\left. m_{2i-1,2j-1}^{(n,k,l)} \right|_{p=p_{0}, \ q=q_{0}} \right),
\]
with $p_0, q_0$ given by Eq. (\ref{p0q0}), would satisfy this dimension reduction condition. Under this reduction, the bilinear equation (\ref{NewKPBilineEq3}) becomes
\begin{eqnarray}
&& \left(\textrm{i}D_{x_{1}}-1 \right) \tau_{n, k, l} \cdot \tau_{n-1, k, l}+ \tau_{n, k-1, l} \tau_{n-1, k+1, l}=0,\label{NewKPBilineEq5}
\end{eqnarray}
and (\ref{NewKPBilineEq4}), combined with (\ref{NewKPBilineEq5}), reduces to
\begin{eqnarray}
D^2_{x_{1}}  \tau_{n, k, l} \cdot \tau_{n-1, k, l}   +  {\rm i}  D_{x_{1}} \tau_{n, k-1, l}  \cdot \tau_{n-1, k+1, l}  =(2  {\rm i} c+1)(  \tau_{n, k-1, l}  \cdot \tau_{n-1, k+1, l} -\tau_{n, k, l} \cdot \tau_{n-1, k, l}).  \label{NewKPBilineEq7}
\end{eqnarray}

\subsection{The index reduction}
The key step to reduce the bilinear equation (\ref{NewKPBilineEq1}) to (\ref{NewdnlsbilinEq1b}) is the observation that the current three-index $\tau$ function (\ref{tausol2}) admits the following index relation,
\[ \label{IndIdentityRe}
\tau_{n, k-1, l} = K^{N} \tau_{n-1, k, l-1},\ \ \ K=\left( \frac{\sqrt{\alpha}+\textrm{i}}{\sqrt{\alpha}-\textrm{i}}\right)^2.
\]
Its proof resembles that in Ref. \cite{OhtaJKY2014} for showing a similar index relation but for a different integrable equation. From the definition of $m_{i,j}^{(n,k,l)} $ in Eq. (\ref{mijnklfunc}), we have
\begin{eqnarray*}
&& m_{i,j}^{(n,k-1,l)}=\mathcal{A}_{i} \mathcal{B}_{j} m^{(n,k-1, l)}=\mathcal{A}_{i} \mathcal{B}_{j} \left(\frac{p}{q}\right) \left(-\frac{q+c}{p-c}\right) \left(\frac{p-d}{q+d}\right)  m^{(n-1,k, l-1)}.
\end{eqnarray*}
Defining
\begin{equation*}
H(\hat{p})=\frac{p(p-d)}{p-c}, \quad  \widetilde{H}(\hat{q})=-\frac{q+c}{q(q+d)},
\end{equation*}
where $\hat{p}= p-c$ and $\hat{q}= q+c$, then
\begin{equation*}
m_{i,j}^{(n,k-1, l)} =\mathcal{A}_{i} \mathcal{B}_{j} H(\hat{p}) \widetilde{H}(\hat{q}) m^{(n-1,k, l-1)}.
\end{equation*}
From the Leibniz rule, we can rewrite the above equation as
\begin{equation*}
m_{i,j}^{(n,k-1, l)}=\sum_{\begin{subarray}{c} \nu=0 \end{subarray}}^{i} \sum_{\begin{subarray}{c} r=0 \end{subarray}}^{j} \frac{1}{\nu !} \frac{1}{r!} H_{\nu}(\hat{p})  \widetilde{H}_r(\hat{q}) \ m_{i-\nu,j-r}^{(n-1,k,l-1)},
\end{equation*}
where functions $H_{\nu}(\hat{p})$ and $\widetilde{H}_r(\hat{q})$ are defined as
\begin{eqnarray*}
&&H_{\nu}(\hat{p})= \left( \hat{p}\partial_{\hat{p}} \right)^\nu  H(\hat{p}), \quad \widetilde{H}_{r}(\hat{q})= \left( \hat{q}\partial_{\hat{q}} \right)^r  \widetilde{H}(\hat{q}).
\end{eqnarray*}
Introducing two generators
\begin{equation*}
\mathcal{G}_{1}=\sum_{\nu=0}^{\infty} \frac{\zeta^\nu}{\nu!} \left( \hat{p}\partial_{\hat{p}} \right)^\nu, \quad
\mathcal{G}_{2}=\sum_{r=0}^{\infty} \frac{\lambda^r}{r!} \left( \hat{q}\partial_{\hat{q}} \right)^r,
\end{equation*}
and using the formula (\ref{Gdef}), we get
\begin{eqnarray*}
&& \mathcal{G}_{1} H(\hat{p}) = H( e^{\zeta} \hat{p})= e^{\zeta}  \hat{p} + \frac{c(c-d)}{\hat{p}}  e^{-\zeta} + 2c-d, \\
&& \mathcal{G}_{2} \widetilde{H}(\hat{q}) = \widetilde{H}( e^{\lambda} \hat{q})= \frac{-1}{e^{\lambda}  \hat{q} +  \frac{c(c-d)}{\hat{q}} e^{-\lambda} -2c+d}.
\end{eqnarray*}
For the chosen $c, d$ values (\ref{cdvalues}) and values $\hat{p}_{0}=\hat{q}_{0}=\sqrt{\alpha}$ from (\ref{p0q0}), we see that $\mathcal{G}_{1} H(\hat{p}_{0})$ and $\mathcal{G}_{2} \widetilde{H}(\hat{q}_{0})$ are even functions of $\zeta$ and $\lambda$, respectively. Thus,
$H_{2\nu-1}(\hat{p}_{0})= \widetilde{H}_{2\nu-1}(\hat{q}_{0})=0$ for all $\nu\ge 1$.
Utilizing these results, we get the relation
\begin{eqnarray*}
m_{i,j}^{(n,k-1,l)}\left|_{p=p_{0}, q=q_{0}} \right.=\sum_{\begin{subarray}{c} \nu=0,\\ \nu: even \end{subarray}}^{i} \sum_{\begin{subarray}{c} r=0,\\ r: even \end{subarray}}^{j} \frac{1}{\nu !} \frac{1}{r!} H_{\nu}(\hat{p})  \widetilde{H}_r(\hat{q}) \ m_{i-\nu,j-r}^{(n-1,k,l-1)}\left|_{p=p_{0}, q=q_{0}} \right..
\end{eqnarray*}
Thus,
\begin{equation*}
\left(\left. m_{2i-1,2j-1}^{(n,k-1,l)} \right|_{p=p_{0}, \ q=q_{0}}\right)_{1\le i,j\le N}=L \left(\left. m_{2i-1,2j-1}^{(n-1,k,l-1)} \right|_{p=p_{0}, \ q=q_{0}}\right)_{1\le i,j\le N}U,
\end{equation*}
where $L$ is a certain lower triangular matrix with $H_{0}( \hat{p}_{0} )$ on the diagonal, and $U$ is a certain upper triangular matrix with $\widetilde{H}_{0}( \hat{q}_{0} )$ on the diagonal.
Taking determinants to this equation, we get
\begin{equation*}
\tau_{n,k-1,l} = \left[H_{0}( \hat{p}_{0} ) \widetilde{H}_{0}( \hat{q}_{0} )  \right]^{N}  \tau_{n-1,k,l-1},
\end{equation*}
which is the same as (\ref{IndIdentityRe}) since $H_0( \hat{p}_{0} ) \ \widetilde{H}_0(\hat{q}_{0}) = K$.

\subsection{Rogue wave solutions} \label{sec:rogue2}
Now, we set $x_1=x-2t$, $x_2={\rm i}t$, $c, d$ as in (\ref{cdvalues}), and $n=k=l=0$ in the above bilinear equations (\ref{NewKPBilineEq1}), (\ref{NewKPBilineEq2}), (\ref{NewKPBilineEq5}) and (\ref{NewKPBilineEq7}). Since $\tau_{0,0,1}=\left( K \right)^{N}\tau_{-1,1,0}$ due to the index relation (\ref{IndIdentityRe}), we find that when we define
\begin{equation*}
f=\tau_{0,0,0}, \quad \bar{f}=\tau_{-1,0,0}, \quad g=\tau_{-1,1,0}, \quad \bar{g}=\tau_{0,-1,0},
\end{equation*}
the above bilinear equations would become
\begin{eqnarray}\label{Bilineformfgh}
\begin{array}{ll}
  \left(\mathrm{i}D_{t}+D_{x}^{2}-2\mathrm{i} \alpha D_{x} \right) g\cdot f=0, \\
\left({\rm i} D_t + D^2_x +2{\rm i}  D_x  \right) f \cdot \bar{f}=0, \\
  ({\rm i}D_{x}-1) f\cdot \bar{f}+g \bar{g}=0,   \\
  D_{x}^2f\cdot \bar{f}-{\rm i}D_x g \cdot \bar{g}+(2\alpha+1)(f\bar{f}-g \bar{g})=0.
\end{array}
\end{eqnarray}
Notice that these $(f, \bar{f}, g, \bar{g})$ functions all have index $l=0$. Thus, these functions are exactly the same as those given in Eq. (\ref{fgdef1}) of the earlier section. Then, following the same complex-conjugacy reductions $\bar{f}=f^*$ and $\bar{g}=g^*$ as before,
the bilinear system (\ref{Bilineformfgh}) reduces to Eqs. (\ref{NewdnlsbilinEq1b})-(\ref{NewdnlsbilinEq4b}), and its rogue wave solutions are exactly as given in Theorems 1 and 3.

\section{Conclusions and Discussions}
In this article, we have derived general rogue waves in the GDNLS equations (\ref{GDNLS0}) by an improved bilinear KP reduction method. Since these GDNLS equations arise in multiple physical situations and contain the Kaup-Newell equation, the Chen-Lee-Liu equation and others as special cases, these results would be useful for rogue-wave generation in such physical systems. A main benefit of this bilinear framework is that, rogue waves to all members of these GDNLS equations can be expressed by the same bilinear solution. Compared to previous bilinear KP reduction methods for rogue waves in other integrable equations, an important improvement in our current KP reduction technique is a new parameterization of internal parameters in rogue waves. Under this new parameterization, the bilinear solution is much simpler than before.
In addition, the rogue wave with the highest peak amplitude at each order can be easily obtained by setting all these internal parameters to zero. This way, the maximum peak amplitude at order $N$ is found to be $2N+1$ times the background amplitude, independent of the individual GDNLS equation and the background wavenumber. We have also found that these GDNLS equations can be decomposed into two different bilinear systems which require different KP reductions, but the resulting rogue waves are the same. Dynamics of rogue waves in the GDNLS equations is also analyzed. It is shown that the wavenumber of the constant background strongly affects the orientation and duration of the rogue wave. In addition, some new rogue patterns are presented.

The GDNLS equations (\ref{GDNLS0}) considered in this article have the parameter requirement of $a\ne b$, in which case these equations are gauge-equivalent to the derivative NLS equation of Kaup-Newell type (\ref{KNE}) (see Sec. \ref{sec:pre}). If $a=b$, Eq. (\ref{GDNLS0}) is called the Kundu-Eckhaus equation in the literature \cite{Kundu1984}. The Kundu-Eckhaus equation is gauge-equivalent to the NLS equation rather than the derivative NLS equation, and thus its rogue waves would be different from those for the GDNLS equations (\ref{GDNLS0}) with $a\ne b$. Rogue waves in the Kundu-Eckhaus equation have been studied by Darboux transformation in \cite{KE_rogue_2013,KE_rogue_2014,KE_rogue_2015}. In the bilinear framework, we can derive general rogue waves in the Kundu-Eckhaus equation in a similar way as we did for the GDNLS equations (\ref{GDNLS0}) with $a\ne b$. This derivation will be sketched in the appendix.

\section*{Acknowledgement}
The work of B.Y. and J.Y. is supported in part by the National Science Foundation (DMS-1910282) and the Air Force Office of Scientific Research (FA9550-18-1-0098), and the work of J.C. is supported by the National Natural Science Foundation of China (No.11705077). J.C. thanks J.Y. and the University of Vermont for hospitality during his visit, where this work was done.

\begin{center}
\textbf{Appendix: Bilinear derivation of rogue waves in the Kundu-Eckhaus equation}
\end{center}

When $a=b$, Eq. (\ref{GDNLS0}) becomes the Kundu-Eckhaus equation \cite{Kundu1984}
\[ \label{KE}
\textrm{i} \phi_{t} + \phi_{\xi\xi} + \rho |\phi|^2\phi + \textrm{i}a (|\phi|^2)_{\xi}\phi  +\frac{1}{4}a^2|\phi|^4\phi=0.
\]
Under a gauge transformation
\begin{equation*}
\phi(\xi,t)=w(\xi,t) e^{-\frac{a}{2}\textrm{i}\int |w(\xi,t)|^2 d\xi},
\end{equation*}
this Kundu-Eckhaus equation reduces to the NLS equation
\[ \label{wNLS}
\textrm{i}w_{t} + w_{\xi\xi} + \rho |w|^2w=0,
\]
whose rogue waves have been derived before \cite{AAS2009,DGKM2010,ACA2010,KAAN2011,GLML2012,OhtaJY2012,DPMVB2013}. To directly obtain rogue waves in the Kundu-Eckhaus equation (\ref{KE}) without the use of the above gauge transformation, we can apply a similar bilinear approach as we did for the $a\ne b$ case in the main text of this article. Specifically, through a scaling of $(\phi, \xi, t, a)$ together with a Galilean transformation, we can normalize $\rho=2$ in Eq. (\ref{KE}), and the boundary conditions of its rogue waves can be normalized as
\[
\phi(\xi, t) \to e^{{\rm i}\left(2t-\frac{1}{2}a\xi\right)}, \quad (\xi, t) \to \infty.
\]
Then, we employ a bilinear variable transformation
\[ \label{bitransformKE}
\phi(\xi,t)=e^{{\rm i}\left[2t-\frac{1}{2}a[\xi+(\ln f)_\xi]\right]}\frac{g}{f},
\]
where $f$ is a real function, and $g$ a complex function. Under this transformation, the Kundu-Eckhaus equation (\ref{KE}) can be split into the following three bilinear equations,
\begin{eqnarray}
&&\left({\rm i} D_t+D_\xi^2\right) g\cdot f=0,  \label{bieqKE1} \\
&& (D_\xi^2+2)f\cdot f=2|g|^2, \label{bieqKE2} \\
&& D_\xi D_tf\cdot f=2{\rm i}D_\xi g\cdot g^*.  \label{bieqKE3}
\end{eqnarray}
One can recognize that the first two bilinear equations are the ones for the NLS equation (\ref{wNLS}) with $\rho=2$ \cite{OhtaJY2012}. It turns out that the $(f, g)$ solutions for rogue waves of the NLS equation also satisfy the third bilinear equation above, and thus rogue waves for the Kundu-Eckhaus equation (\ref{KE}) are given by (\ref{bitransformKE}), where $(f,g)$ are those for the NLS equation (\ref{wNLS}).
The reason for this is that, under the same differential and difference relations of $\tau$ functions listed in Eq. (3.7) of Ref. \cite{OhtaJY2012}, the following three multi-dimensional bilinear equations are satisfied simultaneously,
\begin{eqnarray}
&& (D_{x_1}D_{x_{-1}}-2)\tau_{n}\cdot \tau_n=-2\tau_{n+1}\tau_{n-1}, \\
&& (D_{x_2}-D_{x_1}^2)\tau_{n+1}\cdot \tau_n=0, \\
&& D_{x_{-1}}D_{x_2}\tau_{n}\cdot \tau_n=2D_{x_1}\tau_{n-1}\cdot\tau_{n+1}.   \label{bieqkE3b}
\end{eqnarray}
Thus, with the same dimension reduction and complex conjugacy conditions of the NLS equation \cite{OhtaJY2012}, and setting $x_1=\xi$, $x_2={\rm i} t$, these multi-dimensional bilinear equations reduce to (\ref{bieqKE1})-(\ref{bieqKE3}), and thus the $(f, g)$ solutions for rogue waves of the NLS equation (\ref{wNLS}) are also bilinear solutions for rogue waves of the Kundu-Eckhaus equation (\ref{KE}) under the bilinear variable transformation (\ref{bitransformKE}).


\begin{thebibliography}{10}

\bibitem{Akhmediev_2009}
Akhmediev N, Ankiewicz A and Taki M 2009
Waves that appear from nowhere and disappear without a trace,
Phys. Lett. A 373, 675-678.

\bibitem{Peregrine}
Peregrine D H 1983
Water waves, nonlinear Schrodinger equations and their solutions,
J. Aust. Math. Soc. B 25, 16-43.

\bibitem{Ocean_rogue_review}
Dysthe K, Krogstad H E and M\"uller P 2008
Oceanic Rogue Waves,
Annu. Rev. Fluid Mech. 40, 287-310.

\bibitem{Pelinovsky_book}
Kharif C, Pelinovsky E and Slunyaev A 2009
\emph{Rogue Waves in the Ocean} (Springer, Berlin).

\bibitem{Solli_Nature}
Solli D R, Ropers C, Koonath P and Jalali B 2007
Optical rogue waves, Nature 450, 1054-1057.

\bibitem{Wabnitz_book}
Wabnitz S. (Ed.) 2017
\emph{Nonlinear Guided Wave Optics: A testbed for extreme waves} (IOP Publishing, Bristol, UK).

\bibitem{Tank1}
Chabchoub A, Hoffmann N P and Akhmediev N 2011
Rogue wave observation in a water wave tank,
Phys. Rev. Lett. 106, 204502.

\bibitem{Tank2}
Chabchoub A, Hoffmann N, Onorato M, Slunyaev A, Sergeeva A, Pelinovsky E and Akhmediev N 2012
Observation of a hierarchy of up to fifth-order rogue waves in a water tank,
Phys. Rev. E 86, 056601.

\bibitem{Fiber1}
Kibler B, Fatome J, Finot C, Millot G, Dias F, Genty G, Akhmediev N and Dudley J M 2010
The Peregrine soliton in nonlinear fibre optics,
Nat. Phys. 6, 790-795.

\bibitem{Fiber2}
Frisquet B, Kibler B, Morin P, Baronio F, Conforti M, Millot G and Wabnitz S 2016
Optical dark rogue wave,
Sci. Rep. 6, 20785.

\bibitem{Fiber3}
Baronio F, Frisquet B, Chen S, Millot G, Wabnitz S and Kibler B 2018
Observation of a group of dark rogue waves in a telecommunication optical fiber,
Phys. Rev. A 97, 013852.

\bibitem{AAS2009}
Akhmediev N, Ankiewicz A and Soto-Crespo J M 2009
Rogue waves and rational solutions of the nonlinear Schr\"odinger equation,
Phys. Rev. E 80, 026601.

\bibitem{ACA2010}
Ankiewicz A, Clarkson P A and Akhmediev N 2010
Rogue waves, rational solutions, the patterns of their zeros and integral relations,
J. Phys. A 43, 122002.

\bibitem{DGKM2010}
Dubard P, Gaillard P, Klein C and Matveev V B 2010
On multi-rogue wave solutions of the NLS equation and positon solutions of the KdV equation,
Eur. Phys. J. Spec. Top. 185, 247-258.

\bibitem{KAAN2011}
Kedziora D J, Ankiewicz A and Akhmediev N, 2011
Circular rogue wave clusters,
Phys. Rev. E  84, 056611.

\bibitem{GLML2012}
Guo B L, Ling L M and Liu Q P 2012
Nonlinear Schrodinger equation: generalized Darboux transformation and rogue wave solutions,
Phys. Rev. E 85, 026607.

\bibitem{DPMVB2013}
Dubard P and Matveev V B 2013
Multi-rogue waves solutions: from the NLS to the KP-I equation,
Nonlinearity 26, R93-R125.

\bibitem{OhtaJY2012}
Ohta Y and Yang J 2012
General high-order rogue waves and their dynamics in the nonlinear Schr\"odinger equation,
Proc. R. Soc. Lond. A 468, 1716-1740.

\bibitem{KN_rogue_2011}
Xu S W, He J S and Wang L H 2011
The Darboux transformation of the derivative nonlinear Schr\"odinger equation,
J. Phys. A 44, 305203.

\bibitem{KN_rogue_2013}
Guo B L, Ling L M and Liu Q P 2013
High-order solutions and generalized Darboux transformations of derivative nonlinear Schr\"odinger equations,
Stud. Appl. Math. 130, 317-344.

\bibitem{CCL_rogue_Chow_Grimshaw2014}
Chan H N, Chow K W, Kedziora D J, Grimshaw R H J and Ding E 2014
Rogue wave modes for a derivative nonlinear Schr\"odinger model,
Phys. Rev. E 89, 032914.

\bibitem{CCL_rogue_2017}
Zhang Y S, Guo L J, Chabchoub A and He J S 2017
Higher-order rogue wave dynamics for a derivative nonlinear Schr\"odinger equation,
Rom. J. Phys. 62, 102.

\bibitem{BDCW2012}
Baronio F, Degasperis A, Conforti M and Wabnitz S 2012
Solutions of the vector nonlinear Schr\"odinger equations: evidence for deterministic rogue waves,
Phys. Rev. Lett. 109, 044102.

\bibitem{ManakovDark}
Baronio F, Conforti M, Degasperis A, Lombardo S, Onorato M and Wabnitz S 2014
Vector rogue waves and baseband modulation instability in the defocusing regime,
Phys. Rev. Lett. 113, 034101.

\bibitem{OhtaJKY2012}
Ohta Y and Yang J 2012
Rogue waves in the Davey-Stewartson I equation,
Phys. Rev. E 86, 036604.

\bibitem{OhtaJKY2013}
Ohta Y and Yang J 2013
Dynamics of rogue waves in the Davey-Stewartson II equation,
J. Phys. A 46, 105202.

\bibitem{OhtaJKY2014}
Ohta Y and Yang J 2014
General rogue waves in the focusing and defocusing Ablowitz-Ladik equations,
J. Phys. A 47, 255201.

\bibitem{YangYang2019Nonloc}
Yang B and Yang J 2019
On general rogue waves in the parity-time-symmetric nonlinear Schr\"odinger equation,
arXiv:1903.06203 [nlin.SI].

\bibitem{JCChen2018LS}
Chen J, Chen Y, Feng B F, Maruno K I and Ohta Y 2018
General high-order rogue waves of the (1+1)-dimensional Yajima-Oikawa system,
J. Phys. Soc. Jpn. 87, 094007.

\bibitem{XiaoeYong2018}
Zhang X and Chen Y 2018
General high-order rogue waves to nonlinear Schr\"{o}dinger-Boussinesq equation with the dynamical analysis,
Nonlinear Dyn. 93, 2169-2184.

\bibitem{AANJM2010}
Ankiewicz A, Akhmediev N and Soto-Crespo J M 2010
Discrete rogue waves of the Ablowitz-Ladik and Hirota equations,
Phys. Rev. E 82, 026602.

\bibitem{ASAN2010}
Ankiewicz A, Soto-Crespo J M and Akhmediev N 2010
Rogue waves and rational solutions of the Hirota equation,
Phys. Rev. E 81, 046602.

\bibitem{Chow}
Chow K W, Chan H N, Kedziora D J and Grimshaw R H J 2013
Rogue wave modes for the long wave-short wave resonance model,
J. Phys. Soc. Jpn. 82, 074001.

\bibitem{MuQin2016}
Mu G and Qin Z 2016
Dynamic patterns of high-order rogue waves for Sasa-Satsuma equation,
Nonlinear Anal. Real World Appl. 31, 179-209.

\bibitem{LLMFZ2016}
Ling L M, Feng B F and Zhu Z 2016
Multi-soliton, multi-breather and higher order rogue wave solutions to the complex short pulse equation,
Physica D 327, 13-29.

\bibitem{ClarksonDowie2017}
Clarkson P A and Dowie E 2017
Rational solutions of the Boussinesq equation and applications to rogue waves,
Trans. Math. Appl. 1, 1-26.

\bibitem{Akhmediev_rogue_Raman}
Ankiewicz A, Bokaeeyan M and Akhmediev N 2018
Rogue waves under influence of Raman delay,
J. Opt. Soc. Am. B 35, 899-908.

\bibitem{Kundu1984}
Kundu A 1984
Landau-Lifshitz and higher-order nonlinear systems gauge generated from nonlinear Schr\"odinger-type equations,
J. Math. Phys. 25, 3433-3438.

\bibitem{Clarkson1987}
Clarkson P A and Cosgrove C M 1987
Painlev\'e analysis of the nonlinear Schr\"odinger family of equations,
J. Phys. A 20, 2003-2024.


\bibitem{Agrawal_book}
Agrawal G P 2001
\emph{Nonlinear Fiber Optics }(3rd edition) (Academic Press, San Diego).

\bibitem{Kivshar_book}
Kivshar Y S and Agrawal G P 2003
\emph{Optical Solitons: From Fibers to Photonic Crystals}
(Academic Press, San Diego).

\bibitem{Kaup_Newell}
Kaup D J and Newell A C 1978
An exact solution for a derivative nonlinear Schr\"odinger equation,
J. Math. Phys. 19, 798-801.

\bibitem{KN_Alfven1}
Mio K, Ogino T, Minami K and Takeda S 1976 Modified nonlinear Schr\"{o}inger equation for Alfv\'en waves propagating along the magnetic field in cold plasmas, J. Phys. Soc. Jpn. 41, 265.

\bibitem{KN_Alfven2}
Mjolhus E 1976
On the modulational instability of hydromagnetic waves parallel to the magnetic field,
J. Plasma Phys. 16, 321-334.

\bibitem{CCL}
Chen H H, Lee Y C and Liu C S 1979
Integrability of nonlinear Hamiltonian systems by inverse scattering method,
Phys. Scr. 20, 490.

\bibitem{Wise_CCL}
Moses J, Malomed B A and Wise F W 2007
Self-steepening of ultrashort optical pulses without self-phasemodulation,
Phys. Rev. A 76, 021802.

\bibitem{GI}
Gerdjikov V S and Ivanov I 1983
A quadratic pencil of general type and nonlinear evolution equations. II. Hierarchies of Hamiltonian structures,
Bulg. J. Phys. 10, 130-143.

\bibitem{GI_rogue_2012}
Xu S W and He J S 2012
The rogue wave and breather solution of the Gerdjikov-Ivanov equation,
J. Math. Phys. 53, 063507.

\bibitem{GI_rogue_2014}
Guo L J, Zhang Y S, Xu S W, Wu Z W and He J S 2014
The higher order rogue wave solutions of the Gerdjikov-Ivanov equation,
Phys. Scr. 89, 035501.

\bibitem{GDNLS_rogue2019}
Chen S, Zhou Y, Bu L, Baronio F, Soto-Crespo J M and Mihalache D 2019
Super chirped rogue waves in optical fibers,
Opt. Exp. 27, 11370-11384.

\bibitem{Satsuma_GDNLS_soliton}
Kakei S, Sasa N and Satsuma J 1995
Bilinearization of a generalized derivative nonlinear Schr\"odinger equation,
J. Phys. Soc. Jpn. 64, 1519-1523.

\bibitem{HeNLSheight}
Wang L, Yang C H, Wang J and He J S 2017
The height of an $n$th-order fundamental rogue wave for the nonlinear Schr\"odinger equation,
Phys. Lett. A 381, 1714-1718.

\bibitem{JCChen2018D-LS}
Chen J, Feng B.F, Maruno K and Ohta Y 2018
The derivative Yajima-Oikawa system: bright, dark soliton and breather solutions,
Stud. Appl. Math. 141, 145-185.

\bibitem{BFFengOhta2016CSP}
Feng B F, Maruno K I and Ohta Y 2017
Geometric formulation and multi-dark soliton solution to the defocusing complex short pulse equation,
Stud. Appl. Math. 138, 343-367.

\bibitem{SS_twosets}
Gilson C, Hietarinta J, Nimmo J and Ohta Y 2003
Sasa-Satsuma higher-order nonlinear Schr\"odinger equation and its bilinearization and multisoliton solutions,
Phys. Rev. E 68, 016614.

\bibitem{KE_rogue_2013}
Zhaqilao 2013
On Nth-order rogue wave solution to the generalized nonlinear Schr\"odinger equation,
Phys. Lett. A 377, 855-859.

\bibitem{KE_rogue_2014}
Wang X, Yang B, Chen Y and Yang Y Q 2014
Higher-order rogue wave solutions of the Kundu-Eckhaus equation,
Phys. Scr. 89, 095210.

\bibitem{KE_rogue_2015}
Qiu D Q, He J S, Zhang Y S and Porsezian K 2015
The Darboux transformation of the Kundu-Eckhaus equation,
Proc. R. Soc. A 471, 20150236.



\end{thebibliography}
\end{document}